\documentclass[largeformat]{interact}

\usepackage{epstopdf}
\usepackage[caption=false]{subfig}

\usepackage[numbers,sort&compress,merge]{natbib}
\bibpunct[, ]{[}{]}{,}{n}{,}{,}

\theoremstyle{plain}

\theoremstyle{definition}

\theoremstyle{remark}

\usepackage{ytableau}
\newcommand{\bs}[1]{{\boldsymbol{#1}}}
\newcommand{\cdag}[2]{{c_{#1 #2}^\dagger}}
\newcommand{\anh}[2]{{c_{#1 #2}^{\phantom{\dagger}}}}
\newcommand{\pdag}[2]{{p_{#1 #2}^\dagger}}
\newcommand{\gdag}[2]{{\gamma_{#1 #2}^\dagger}}

\newcommand{\ket}[1]{|#1\rangle}
\newcommand{\bra}[1]{\langle #1 \lvert}
\newcommand{\braket}[2]{\langle #1 \mid #2 \rangle}

\begin{document}


\title{Irreducible Representations as Multireference Indicators for Diradicaloid Systems}

\author{
\name{Emmalyn~A. Sarver\textsuperscript{a} and Lukas Muechler\textsuperscript{a,b}\thanks{CONTACT Lukas Muechler. Email: lfm@psu.edu}}
\affil{\textsuperscript{a}Department of Chemistry, Pennsylvania State University, University Park, Pennsylvania 16802, USA; \textsuperscript{b}Department of Physics, Pennsylvania State University, University Park, Pennsylvania 16802, USA}
}

\maketitle

\begin{abstract}
Multireference behavior in molecules often arises when a small gap between frontier orbitals results in mixing of closed and open-shell configurations. 
Standard multireference diagnostics of this regime usually rely on correlated wavefunctions, natural-orbital occupations, or reduced density matrices. 
Here, we examine a complementary, symmetry-based criterion for a model system. 
For a time-reversal-invariant Hamiltonian, a symmetry-preserving, closed-shell Slater determinant must transform as the trivial irreducible representation of its point group. 
Therefore, a nontrivial, many-electron irreducible representation excludes such a description.
We compare two pathways within the same model to demonstrate this.
Along the control pathway, the frontier orbitals remain separated and the ground state retains a trivial irreducible representation over the weak-to-intermediate interaction regime. 
Along the obstructed pathway, a high-symmetry point produces a frontier-orbital degeneracy, resulting in a singlet ground state with two-configuration character and a nontrivial irreducible representation. 
Exact diagonalization, a two-state effective model, and the Frobenius norm of the two-particle cumulant provide a consistent picture in this regime, demonstrating that irreducible representations can serve as a low-cost diagnostic of multireference character in diradicaloid models. 
While symmetry is not a quantitative measure of correlation strength, it does offer a computationally inexpensive screening tool to identify obstructions to a single-reference description.
\end{abstract}


\section{Introduction}

Multireference character arises when an electronic state is not adequately described by a single Slater determinant. 
This frequently occurs in diradicals, bond dissociations, excited states, and high-symmetry structures with small or vanishing gaps between frontier orbitals~\cite{Lischka2018-vz,Vitillo2022-cd,Park2020-sz,Wardzala2026-bh,Abe2013-io, Stuyver2019-ut}. 
A central issue is often not only the magnitude of electron correlation, but whether a closed-shell, single-determinant description is compatible with the many-electron state.
The question of multireference character is usually addressed with numerical diagnostics extracted from high-level approximations of wavefunctions or reduced density matrices. 
For example, natural-orbital occupations, configuration weights, and cumulant-based measures are each frequently used to gauge the multireference character of a molecule~\cite{Ganoe2024-rm, Xu2025-zt,Coe2015-jy}. 
These tools are essential to diagnose multireference correlations, but they require a converged, high-level approximation of the true wavefunction to be predictive. 

Because multireference methods are substantially more computationally demanding than single-reference methods, determining whether a state admits a \mbox{single-reference} description is an important practical question in computational chemistry.
In particular, this question is relevant since a molecule's multireference character may change as a result of changes in its geometry. For example, cyclobutadiene in its $D_{2h}$ geometry is amendable to a single-reference description. However, its $D_{4h}$ geometry is not compatible with single-reference methods and requires multireference or spin-flip approaches. Related, symmetry-driven, open-shell behavior also appears in $\pi$-conjugated hydrocarbons, Jahn-Teller active molecules, and symmetry-forbidden reactions~\cite{Ortiz2019-ov,Muechler-Mobius,Mirzanejad2025-fj}.

The relevant connection between symmetry and this obstruction follows from an obstruction argument introduced by Yao and Kivelson~\cite{Yao-FMI}. 
For a time-reversal-invariant Hamiltonian, any closed-shell Slater determinant built from Kramers doublets transforms as the identity representation under point-group operations that commute with time-reversal symmetry (TRS). Consequently, a correlated many-electron state with a nontrivial irreducible representation (irrep) cannot be represented by a  time-reversal-invariant, closed-shell, single determinant. The irrep does not measure the magnitude of correlation, but does provide a rigorous answer to the question regarding whether a state is amendable to single-reference descriptions.
This connection between symmetries and multireference character is further crucial in context of the development of symmetry-projected methods~\cite{Scuseria2011-kz,Jimenez-Hoyos2012-wv,Song2024-ap,Izsak2023-ln,Song2025-gr}.

The purpose of the present work is to examine this connection for a model designed to isolate the change between single-reference and multireference character along reaction pathways, and to compare it to conventional measures of strong correlations.
We study a six-site Hubbard Hamiltonian containing three coupled dimers and compare two pathways parameterized by a single coordinate ($\alpha$). The first pathway serves as a control case in which the frontier orbitals remain separated. The second features a crossing of the highest occupied molecular orbital (HOMO) and the lowest unoccupied molecular orbital (LUMO) of different symmetries. This construction allows us to ask, within the same interacting Hamiltonian, how the many-electron irrep changes as the ground state evolves from a predominantly closed-shell state to a two-configuration, diradicaloid state.

\section{Conventions and Model Setup}
\label{sec:Set-Up}

We consider a six-site Hubbard model which describes three dimers with intra-dimer hopping ($1-\alpha$) and inter-dimer hopping $\alpha$, where we assume $\alpha \equiv \alpha(t)$ to be time-dependent and we treat it as an adiabatic reaction coordinate parameter.
The model describes an unobstructed, control pathway, along which there are no crossings of the frontier orbitals, and a symmetry-obstructed pathway, which features a crossing of the HOMO and the LUMO.

The Hamiltonian describing these models in an atomic-orbital basis is written as
\begin{equation}
    \begin{split}
        H(\alpha) &= \sum_{i,j=1}^6 \sum_{\sigma=\uparrow,\downarrow} h_{ij}(\alpha)
        \cdag{i}{\sigma} \anh{j}{\sigma} + U \sum_{j=1}^6 \cdag{j}{\uparrow} \anh{j}{\uparrow} \cdag{j}{\downarrow} \anh{j}{\downarrow} \\
        & + \mu \sum_{j=1}^6 
        (\cdag{j}{\uparrow} \anh{j}{\uparrow} + \cdag{j}{\downarrow} \anh{j}{\downarrow})
        \end{split}
\end{equation}
where $\boldsymbol{h}$ is the single-particle hopping matrix, $U$ is the on-site interaction strength, and $\mu$ is the chemical potential, which is used to control the number of electrons.
The single-particle Hamiltonian $\boldsymbol{h}_{\beta}(\alpha)$ is used to describe both pathways,
\begin{equation}
   \boldsymbol{h}_{\beta}(\alpha)=-
\begin{bmatrix}
 0 & 1-\alpha  & \alpha  h & 0 & \beta  h & \beta  \\
 1-\alpha  & 0 & \alpha  & \alpha  h & 0 & \beta  h \\
 \alpha  h & \alpha  & 0 & 1-\alpha  & \alpha  h & 0 \\
 0 & \alpha  h & 1-\alpha  & 0 & \alpha  & \alpha  h \\
 \beta  h & 0 & \alpha  h & \alpha  & 0 & 1-\alpha  \\
 \beta  & \beta  h & 0 & \alpha  h & 1-\alpha  & 0 \\
\end{bmatrix}
\end{equation}
where $h$ is a small, constant parameter modeling the next-nearest neighbor hopping.
The two pathways are distinguished by the hopping between the two terminal sites,  $h_{16} = h_{61} = \beta $. 
The unobstructed, control pathway corresponds to $\beta = \alpha$, and the obstructed pathway corresponds to $\beta = - \alpha$. 
The obstructed pathway is denoted by $\boldsymbol{h_-}(\alpha)$ and the unobstructed pathway is denoted by $\boldsymbol{h_+}(\alpha)$.
Chemically, $\alpha$ correlates with the degree of electron delocalization between overlapping orbitals, i.e., the extent of bond formation. Here, it is used as a reaction coordinate, where $\alpha = 0$ ($\alpha = 1$) corresponds to the reactant (product).
This model is not a quantitative model for reactions, but rather a minimal, symmetry-constrained example, allowing us to describe both pathways with the same set of parameters, sites, and interactions. 

To determine the irreps along each pathway, we first specify the relevant point groups and the matrix representations of their generators. Along the unobstructed pathway, the structures have $D_{3h}$ symmetry, except at the higher-symmetry point $\alpha=\frac{1}{2}$ where the symmetry is $D_{6h}$.
For $D_{3h}$, we use $C_3$, $C_2'$, and $\sigma_h$ as generators. 
For $D_{6h}$, we use $C_6$, $C_2''$, and $I$ as generators. The matrix representations of these generators in the atomic-orbital basis are given in Table~\ref{tab:symmetryMatrices}. 
For this model, the $D_{3h}$ symmetry $C_2'$ is equivalent to the $D_{6h}$ symmetry $C_2''$, and both are referred to as $C_2''$ in this manuscript.

\begin{table}
    \tbl{Matrix representations of point group generators along both unobstructed and obstructed pathways in the atomic-orbital basis, where $\Bar{1} = -1$. Here, operators $C_6$, $C_2''$, and $I$ are generators of $D_6h$; operators $C_3$, $C_2''$, and $\sigma_h$ are generators of $D_3h$; $C_{3,o}$ and $\sigma_{d,o}$ are generators of $\tilde{C}_{3v}$; and $C_{6,o}$ and $\sigma_{d,o}$ are generators of $\tilde{C}_{6v}$. The subscript $o$ is used for the elements of the double groups to distinguish them from their single point group counterparts, where $o$ is used for the obstructed pathway.} 
    {\begin{tabular}{cc|ccc|ccc|cc}
     $D(C_6)= $&  &   & $D(C_3)= $& & & $D\left(C_2''\right)= $ & & & $D(\sigma_h) $ \\ 
 $\begin{bmatrix}
 0 & 0 & 0 & 0 & 0 & 1 \\
 1 & 0 & 0 & 0 & 0 & 0 \\
 0 & 1 & 0 & 0 & 0 & 0 \\
 0 & 0 & 1 & 0 & 0 & 0 \\
 0 & 0 & 0 & 1 & 0 & 0 \\
 0 & 0 & 0 & 0 & 1 & 0 \\
\end{bmatrix}
$ & & &
$
\begin{bmatrix}
 0 & 0 & 0 & 0 & 1 & 0 \\
 0 & 0 & 0 & 0 & 0 & 1 \\
 1 & 0 & 0 & 0 & 0 & 0 \\
 0 & 1 & 0 & 0 & 0 & 0 \\
 0 & 0 & 1 & 0 & 0 & 0 \\
 0 & 0 & 0 & 1 & 0 & 0 \\
\end{bmatrix}$& & &
  $\begin{bmatrix}
 0 & 0 & 0 & 0 & 0 & \Bar{1} \\
 0 & 0 & 0 & 0 & \Bar{1} & 0 \\
 0 & 0 & 0 & \Bar{1} & 0 & 0 \\
 0 & 0 & \Bar{1} & 0 & 0 & 0 \\
 0 & \Bar{1} & 0 & 0 & 0 & 0 \\
 \Bar{1} & 0 & 0 & 0 & 0 & 0 \\
\end{bmatrix}$
& & & $
\begin{bmatrix}
 \Bar{1} & 0 & 0 & 0 & 0 & 0 \\
 0 & \Bar{1} & 0 & 0 & 0 & 0 \\
 0 & 0 & \Bar{1} & 0 & 0 & 0 \\
 0 & 0 & 0 & \Bar{1} & 0 & 0 \\
 0 & 0 & 0 & 0 & \Bar{1} & 0 \\
 0 & 0 & 0 & 0 & 0 & \Bar{1} \\
\end{bmatrix}$ \\
& & & & & & & & &\\ \hline
& & & & & & & & & \\ 
 $D(I)=$ &  & & $D(C_{3,o})=$ & & & $D(C_{6,o})=$ & & & $D(\sigma_{d,o})=$ \\ 
$
\begin{bmatrix}
 0 & 0 & 0 & \Bar{1} & 0 & 0 \\
 0 & 0 & 0 & 0 & \Bar{1} & 0 \\
 0 & 0 & 0 & 0 & 0 & \Bar{1} \\
 \Bar{1} & 0 & 0 & 0 & 0 & 0 \\
 0 & \Bar{1} & 0 & 0 & 0 & 0 \\
 0 & 0 & \Bar{1} & 0 & 0 & 0 \\
\end{bmatrix}$ & & & 
$
\begin{bmatrix}
    0 & 0 & 0 & 0 & \Bar{1} & 0 \\
    0 & 0 & 0 & 0 & 0 & \Bar{1} \\
    1 & 0 & 0 & 0 & 0 & 0 \\
    0 & 1 & 0 & 0 & 0 & 0 \\
    0 & 0 & 1 & 0 & 0 & 0 \\
    0 & 0 & 0 & 1 & 0 & 0 \\ 
\end{bmatrix}$
& & &
    $\begin{bmatrix}
 0 & 0 & 0 & 0 & 0 & \Bar{1} \\
 1 & 0 & 0 & 0 & 0 & 0 \\
 0 & 1 & 0 & 0 & 0 & 0 \\
 0 & 0 & 1 & 0 & 0 & 0 \\
 0 & 0 & 0 & 1 & 0 & 0 \\
 0 & 0 & 0 & 0 & 1 & 0 \\
\end{bmatrix}$
& & &
$
\begin{bmatrix}
 0 & 0 & 0 & 0 & 0 & i \\
 0 & 0 & 0 & 0 & i & 0 \\
 0 & 0 & 0 & i & 0 & 0 \\
 0 & 0 & i & 0 & 0 & 0 \\
 0 & i & 0 & 0 & 0 & 0 \\
 i & 0 & 0 & 0 & 0 & 0 \\
\end{bmatrix}$
\\
    \end{tabular}}
    \label{tab:symmetryMatrices}
\end{table}

The symmetries of the unobstructed pathway do not commute with the Hamiltonian of the obstructed pathway due to the sign-altered hopping introduced between the terminal atoms. 
Instead, we define modified operators $C_{3,o}$ and $C_{6,o}$ which restore the symmetry by taking into account this sign change.
As a result, we now find that  $(C_{3,o})^3= -E \equiv \tilde{E}$ and $(C_{6,o})^6=\tilde{E}$. 
The obstructed pathway is therefore described by double-groups rather than ordinary, single-groups, which are used for the unobstructed pathway~\cite{Altmann1994-ti}.
On the obstructed pathway, the high-symmetry point at $\alpha=\frac{1}{2}$ belongs to the double-group $\tilde{C}_{6v}$, which has generators $C_{6,o}$ and $(\sigma_{d,o})^2=\tilde{E}$. 
Elsewhere on the pathway, the model belongs to the double-group $\tilde{C}_{3v}$, which has generators $C_{3,o}$ and $(\sigma_{d,o})$~\cite{Altmann1994-ti, Dresselhaus2007-yu}. 
The character tables and multiplication tables for these double-groups are contained in the Appendix.

\section{Symmetry Criterion} 
\label{sec:SymmetryCriterion}

We consider a time-reversal symmetric Hamiltonian that transforms according to a point-group $\boldsymbol{G}$ with symmetry elements $\boldsymbol{C}$.
Consider a closed-shell eigenstate of this Hamiltonian that is well approximated as a product state in a single-particle basis:
\begin{equation}
\label{eq:CS}
    \ket{CS} = \prod_{n,\tilde{n} \in occ} \gdag{n}{} \gdag{\tilde{n}}{} \ket{0},
\end{equation}
where $\ket{0}$ is the vacuum state with no electrons, $\gdag{n}{}$ creates an electron in the molecular spin-orbital $\ket{n}$, and ``$occ$ '' is the set of occupied states.  
The states $\ket{n}$ and $\ket{\tilde{n}}$ are linearly independent Kramers doublets, together Kramers partners, that are related by TRS:
\begin{equation}
    T\ket{n} \equiv \ket{\tilde{n}} \text{ and } \braket{n|\tilde{n}} = 0 
\end{equation}
TRS is an anti-unitary operator that is canonically defined as $T = i \sigma_y K$, where $\sigma_y$ is the second Pauli matrix, K is the complex conjugation operator, and $T^2 = -1$. 
Therefore, $\ket{CS}$ is composed of partners of Kramers doublets, each of which are either doubly occupied or unoccupied. 
In the absence of spin-orbit coupling, the Kramers partners are conventionally chosen as spin-up and spin-down electrons residing in the same molecular orbital $n$, i.e., 
\begin{equation}
    \begin{split}
    \ket{n} = & \gdag{n}{\uparrow} \ket{0}, \\
    \ket{\tilde{n}} = & \gdag{n}{\downarrow} \ket{0}.\\ 
    \end{split}
\end{equation}

A single-particle state can always be chosen to be an eigenstate of the symmetries of the point group, i.e., 
\begin{equation}
    C\ket{n} = \lambda_n \ket{n}.
\end{equation}
If $C^m = 1$ for $m \geq 1$, then the eigenvalues of $C$ are 
\begin{equation}
    \lambda_n = e^{i2\pi j_n /m}, \hspace{1cm} j_n = 1, \ldots, m
\end{equation}
For symmetries commuting with TRS, the eigenvalues of the Kramers partners ($\ket{n}$ and $\ket{\tilde{n}}$) under $C$ are related by complex conjugation, $\lambda_n^* = \lambda_{\tilde{n}}$, since 
\begin{equation}
    \begin{split}
        C\ket{\tilde{n}} & = CT\ket{n} \\
        & = TC\ket{n} \\
        & = T\lambda_n \ket{n} \\
        & = T\lambda_n T^{-1} T \ket{n} \\
        & = \lambda_n^* T \ket{n} \\
        & = \lambda_n^* \ket{\tilde{n}}.
    \end{split}
\end{equation}
Thus, the character of a closed-shell state under the symmetry $C$ is constrained,
\begin{equation}
    \begin{split}
        C\ket{CS} & =  \prod_{n,\tilde{n} \in occ} C \  \gdag{n}{} \gdag{\tilde{n}}{} \ket{0} \\
        & = \prod_{n,\tilde{n} \in occ} \lambda_n \lambda_{\tilde{n}} \ket{CS} \\
        & = \prod_{n,\tilde{n} \in occ} \lambda_n \lambda_{n}^* \ket{CS} \\
        & = \prod_{n,\tilde{n} \in occ} \| \lambda_n\|   \ket{CS}  = \ket{CS}.\\
    \end{split}
\end{equation}

Accordingly, a closed-shell state, as defined in Eq.~\ref{eq:CS}, transforms as the trivial irrep of the point group.
This implies that a state not transforming as a trivial representation cannot be expressed as a closed-shell state described by a single Slater-determinant, i.e. it is necessarily multireference.
\begin{equation}
    C\ket{MR} = \lambda \ket{MR} \hspace{1cm} \lambda \neq 1.
\end{equation}
In practice, the same obstruction excludes symmetry-preserving, closed-shell, single-reference descriptions built perturbatively from such a determinant. 
However, it does not follow that a trivial irrep implies weak correlation or guarantees the validity of a single-reference treatment.

\section{Unobstructed Pathway}
\subsection{Single-Particle Properties}
\label{subsec:Allowed-SPprops}

The single-particle hopping matrix along the unobstructed pathway is block-diagonalized into the three eigensectors of its $C_3$ rotation, which is an element of both $D_{3h}$ and $D_{6h}$.
The eigenvalues of $C_3$ are $\lambda_k=exp[ik]$ with $k=0,\pm \frac{2\pi}{3}$. 
The orthogonal transformation into the eigenbasis of $C_3$ is given as,
\begin{equation}
    \bs{K}_{C_3}=\frac{1}{\sqrt{3}}\left[
\begin{array}{cccccc}
 0 & 1 & 0 & e^{-i\frac{2\pi}{3}} & 0 & e^{i\frac{2\pi}{3}}  \\
 1 & 0 & e^{-i\frac{2\pi}{3}} & 0 & e^{i\frac{2\pi}{3}} & 0 \\
 0 & 1 & 0 & e^{i\frac{2\pi}{3}} & 0 & e^{-i\frac{2\pi}{3}} \\
 1 & 0 & e^{i\frac{2\pi}{3}} & 0 & e^{-i\frac{2\pi}{3}} & 0 \\
 0 & 1 & 0 & 1 & 0 & 1 \\
 1 & 0 & 1 & 0 & 1 & 0 \\
\end{array}
\right],
\end{equation}
and the transformed Hamiltonian is
\begin{equation}
    \begin{split}
        \boldsymbol{h}_{C_3}(\alpha)&= \bs{K}_{C_3}^{\dagger}\bs{h}_+(\alpha) \bs{K}_{C_3}\\
        &=
    \left[
    \begin{array}{ccc}
    \bs{h}_0(\alpha) & 0 & 0 \\
    0 & \bs{h}_{\frac{2\pi}{3}}(\alpha) & 0 \\
    0 & 0 & \bs{h}_{-\frac{2\pi}{3}}(\alpha)\\
    \end{array}
    \right],
    \end{split}
\end{equation}
where 
\begin{equation}
\begin{split}
    \bs{h}_0(\alpha) & = \left[
    \begin{array}{cc}
     -2 \alpha h & -1 \\
     -1 & -2 \alpha h \\
    \end{array}
    \right], \\
        \bs{h}_{\frac{2\pi}{3}}(\alpha) & = \left[
    \begin{array}{cc}
     \alpha  h & -1+\frac{3\alpha}{2} + \frac{i\sqrt{3}\alpha}{2}  \\
     -1+\frac{3\alpha}{2} - \frac{i\sqrt{3}\alpha}{2}  & \alpha  h \\
    \end{array}
    \right], \\ 
     \bs{h}_{-\frac{2\pi}{3}}(\alpha)& = \bs{h}_{\frac{2\pi}{3}}(\alpha)^{\ast}.
\end{split}
\end{equation}

Under $\bs{K}_{C_3}$, the creation operators $\cdag{i}{\sigma}$ in the site basis transform as
\begin{equation}
    {\pdag{i,}{\sigma}} = \sum_{j=1}^{6} (K_{C_3})_{ji} \cdag{j}{\sigma}
\end{equation}
leading to a new set of operators,
\begin{equation}
    \begin{split}
        {\pdag{[1,k],}{\sigma}} &= \frac{1}{\sqrt{3}} \sum_{j=1}^{3} e^{-ijk} \cdag{2j,}{\sigma} \\
        {\pdag{[2,k],}{\sigma}} &= \frac{1}{\sqrt{3}} \sum_{j=1}^{3} e^{-ijk} \cdag{2j-1,}{\sigma},
    \end{split}
\end{equation}
where $k$ refers to the eigenvalues $\lambda_k$ of $C_3$.

The block-diagonal, single-particle Hamiltonian $\bs{h}_{C_3}(\alpha)$ can then be fully diagonalized via a unitary transformation by the matrix of its eigenvectors, 
\begin{equation}
    \bs{V}=\frac{1}{\sqrt{2}}\left[
\begin{array}{cccccc}
 1 & -1 & 0 & 0 & 0 & 0 \\
 1 & 1 & 0 & 0 & 0 & 0 \\
 0 & 0 & \frac{x(\alpha)}{y(\alpha)} & -\frac{x(\alpha)}{y(\alpha)} & 0 & 0 \\
 0 & 0 & 1 & 1 & 0 & 0 \\
 0 & 0 & 0 & 0 & \frac{x(\alpha)}{z(\alpha)} & -\frac{x(\alpha)}{z(\alpha)} \\
 0 & 0 & 0 & 0 & 1 & 1 \\
\end{array}
\right] 
\end{equation}
where
\begin{equation} 
    \begin{split}
        x(\alpha) &= 2 i \sqrt{3 \alpha ^2-3 \alpha +1}, \\
        y(\alpha) &= -3 i \alpha -\sqrt{3} \alpha +2 i, \\
        z(\alpha) &= -3 i \alpha +\sqrt{3} \alpha +2 i.
    \end{split}
\end{equation}
This yields the single-particle, MO energies $\epsilon_1^k(\alpha)$, $\epsilon_2^k(\alpha)$
\begin{equation} 
    \begin{split}
        \epsilon_1^0(\alpha) &= -1 - 2 \alpha h \\
        \epsilon_2^0(\alpha) &= 1 - 2 \alpha h \\
        \epsilon_1^{-2\pi/3}(\alpha) = \epsilon_1^{2\pi/3}(\alpha) &=-\sqrt{3 \alpha^2-3 \alpha+1}+\alpha h \\
        \epsilon_2^{-2\pi/3}(\alpha)=\epsilon_2^{2\pi/3}(\alpha) &= \sqrt{3 \alpha^2-3 \alpha+1}+\alpha h
    \end{split}
\end{equation}
and a new set of MO-creation operators, $\gdag{1,k,}{\sigma}$ and $\gdag{2,k,}{\sigma}$, generated by 
\begin{equation}
    \begin{split}
        \gdag{1,k,}{\sigma} &= \frac{1}{\sqrt{2}}(r(k,\alpha)\pdag{[1,k],}{\sigma} + \pdag{[2,k],}{\sigma}) \\
        \gdag{2,k,}{\sigma} &= \frac{1}{\sqrt{2}}(-r(k,\alpha)\pdag{[1,k],}{\sigma} + \pdag{[2,k],}{\sigma}),
    \end{split}
\end{equation}
where
\begin{equation}
    \begin{split}
        r(k,\alpha)&=C(k)\\
        &+\left[1-C(k)\right]\left[\frac{x(\alpha)}{y(\alpha)}\frac{1+S(k)}{2}+\frac{x(\alpha)}{z(\alpha)}\frac{1-S(k)}{2}\right], \\
        C(k) &= \frac{1+2\cos{k}}{3}, \\
        S(k) &= \frac{2}{\sqrt{3}}\sin{k}.
    \end{split}
\end{equation}

\begin{figure}
    \centering
    \resizebox*{10cm}{!}{\includegraphics{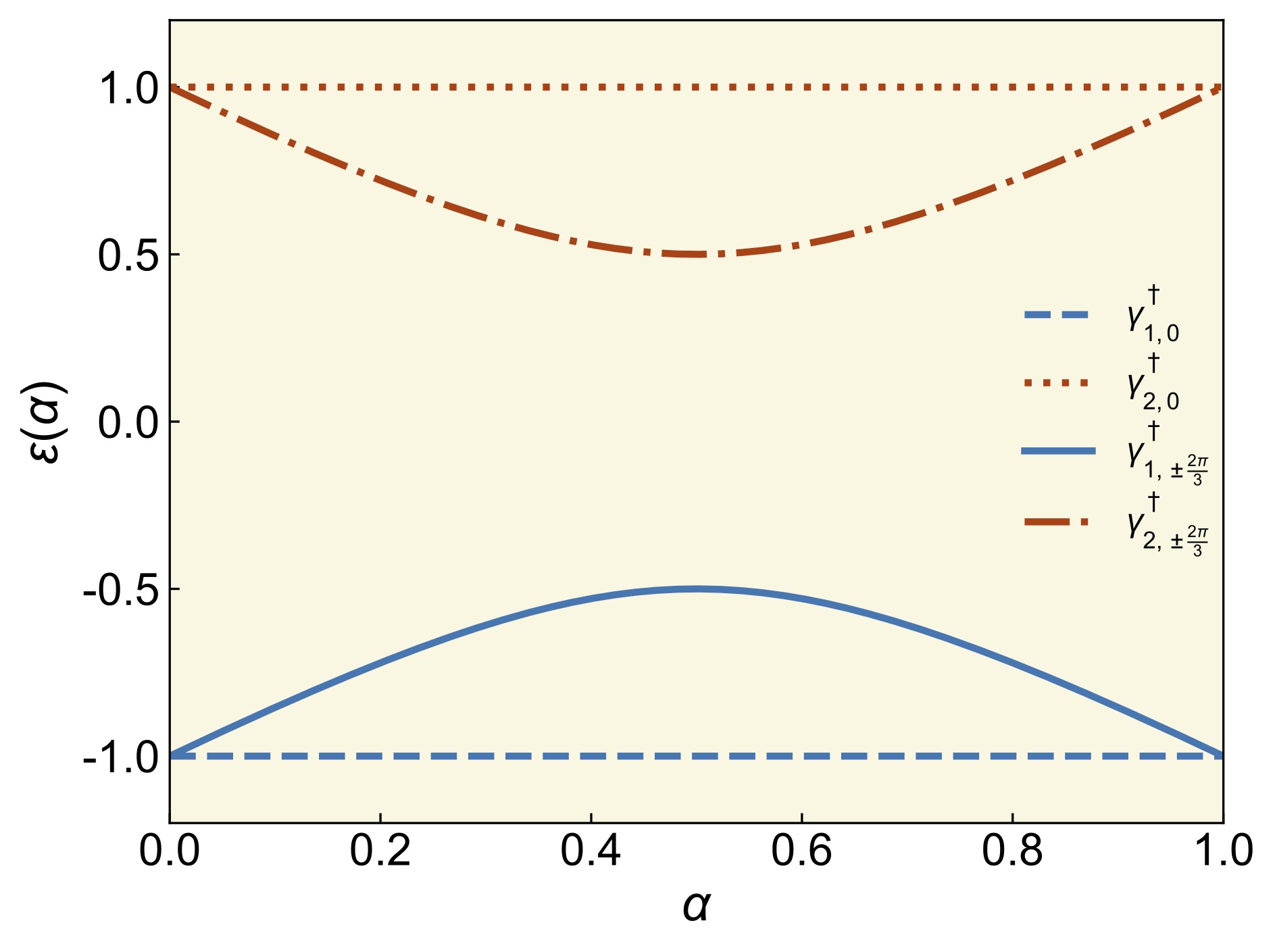}}
    \caption{Single-particle energies of molecular orbitals along the allowed pathway, where $h=0$. There is no crossing of the frontier orbitals.}
    \label{fig:allowedMOenergies}
\end{figure}

Figure~\ref{fig:allowedMOenergies} displays the MO energies for the unobstructed pathway as a function of $\alpha$ for $h = 0$. The unobstructed pathway is characterized by a gap $\Delta_{\alpha}$ between three doubly-occupied MOs and three unoccupied MOs, which is maximal for $\alpha = 0,1$ and takes on its minimal value at $\alpha = \frac{1}{2}$.
The lowest MO ($\gdag{1,}{0}$) is nondegenerate, while the next two MOs are doubly degenerate ($\gdag{1,}{\pm\frac{2\pi}{3}}$).

\subsection{Many-Electron States}
\label{subsec:AllowedMBstates}

\begin{figure}[t]
    \centering
    \resizebox*{10cm}{!}{\includegraphics{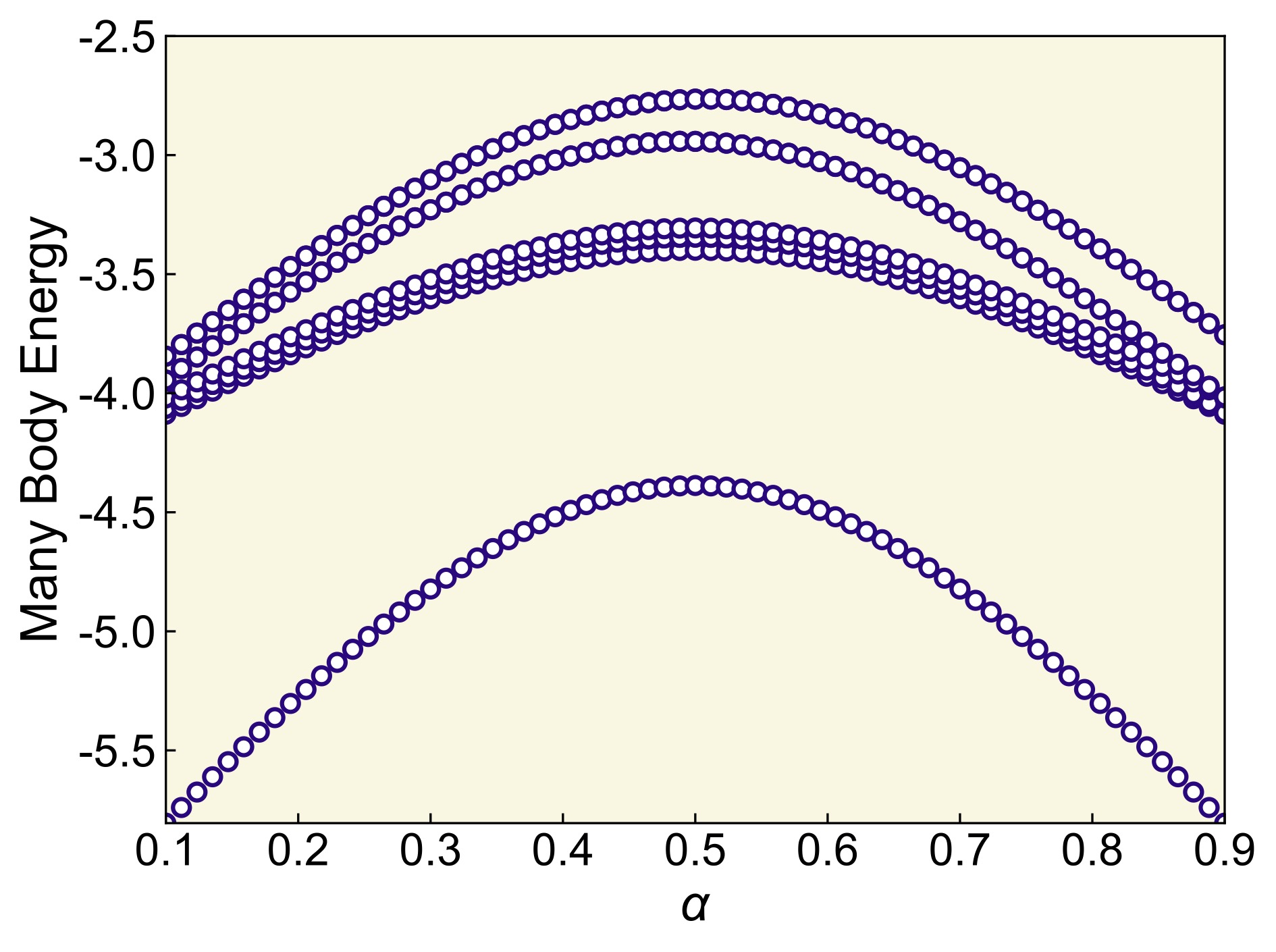}}
    \caption{Many-body energies of the ten lowest singlet states along the allowed pathway, where $h=0.05$, $U=\frac{1}{4}$, and $\mu = -\frac{1}{8}$.}
    \label{fig:allowedSingletMBenergies}
\end{figure}

For $U \ll \Delta_{\alpha}$, the ground state of the interacting Hamiltonian for $U \neq 0$ is well described by a closed-shell product state,
\begin{equation}
\label{eq:gs_approx_allowed}
   \ket{\Psi}= \gdag{1,0,}{\uparrow} \gdag{1,-\frac{2\pi}{3},}{\uparrow} \gdag{1,\frac{2\pi}{3},}{\uparrow} \gdag{1,0,}{\downarrow} \gdag{1,-\frac{2\pi}{3},}{\downarrow} \gdag{1,\frac{2\pi}{3},}{\downarrow} \ket{0},
\end{equation}
where each $\gdag{}{}$ is a function of $\alpha$. 
To verify that Eq.~\ref{eq:gs_approx_allowed} is the true ground state of the model, the Hamiltonian of the unobstructed pathway was solved via exact diagonalization with parameters $h=0.05$, $U=\frac{1}{4}$, and $\mu=-\frac{1}{8}$. 
Overlap of the ground state from exact diagonalization with the product state in Eq.~\ref{eq:gs_approx_allowed} shows that this approximation is well justified up to $U = 1$, for which the overlap is greater than $0.9$.
For larger values of $U$, determinants corresponding to excitations beyond the HOMO/LUMO gap become important and can no longer be neglected. 
The ground state is separated by a large gap from the higher excited singlet states for all values of $\alpha$ [Fig.~\ref{fig:allowedSingletMBenergies}].

\subsection{Irreducible Representations}
\label{subsec:AllowedIrreps}
We now turn to the group-theoretical description of the many-electron states, following the connection between many-electron irreps and multireference character outlined in Sec.~\ref{sec:SymmetryCriterion}. 
Throughout this work, uppercase labels denote irreps of many-electron states, whereas lowercase labels denote irreps of MOs.
The irrep of an MO $\ket{\psi}$ is identified by its characters under each of the generating symmetries $G$ of the molecule's point group.  
\begin{equation}
    \chi (G) = \bra{\psi} G \ket{\psi}
\end{equation}
If two or more orbitals are degenerate, then $\ket{\psi}$ is the matrix of these degenerate orbitals and the character is the trace of $\chi(G)$. 

We evaluate the irreps at the point of highest symmetry, $\alpha = \frac{1}{2}$, and use subduction to determine the irreps for other values of $\alpha$ with a lower symmetry.
Table~\ref{tab:MOIrrepsAllowedAtHalf} shows the characters of the unobstructed pathway's MOs at $\alpha = \frac{1}{2}$ under $D_{6h}$.
\begin{table}[t]
    \tbl{Irreducible representations of the molecular orbitals for the unobstructed pathway at $\alpha=\frac{1}{2}$.}
    {\begin{tabular}{r|cccc|c}
        $D_{6h}$& $E$& $2C_6$& $3C_2''$& $i$& $\Gamma$ \\
        \hline
        ${\gdag{1,0,}{\sigma}}$& 1& 1& -1& -1& $a_{2u}$\\
        ${\gdag{1,\pm \frac{2\pi}{3},}{\sigma}}$& 2& 1& 0& 2& $e_{1g}$\\
        ${\gdag{2,\pm \frac{2\pi}{3},}{\sigma}}$& 2& -1& 0& -2&$e_{2u}$\\
        ${\gdag{2,0,}{\sigma}}$& 1& -1& 1& 1&$b_{2g}$\\
    \end{tabular}}
    \label{tab:MOIrrepsAllowedAtHalf}
\end{table}
Each of the electron shells of the ground state are either fully occupied or unoccupied, and thus the ground state transforms as $A_{1g}$~\cite{Ellis1971-pk},
\begin{equation}
        \Gamma_{\text{Ground State}} = \Gamma_{\text{core}} \otimes  \Gamma_{\text{valence}} = a_{1g} \otimes a_{1g} = A_{1g}.
\end{equation}

Using subduction, the ground state irrep for $\alpha \not= \frac{1}{2}$ is $A_1^{'}$ under $D_{3h}$.
Therefore, this ground state on the unobstructed pathway always transforms as the trivial irrep of its point group.

We now consider the irreps of the lowest excited states at $\alpha = \frac{1}{2}$ which are the result of a single electronic excitation, such that there are two electrons in $a_{2u}$, three electrons in $e_{1g}$, and one electron in $e_{2u}$.
Therefore, there are two electron shells which are partially unoccupied and do not transform as the trivial representation of the point group. 
We investigate these states as a control case for an open-shelled system.

Utilizing the $D_6$ character table and following Ford's method outlined in Ref.~\cite{Ford1972-ug}, we find that the three-electron shell transforms as $e_{1g}$ (See~\ref{sec:Appendix} for details).
A singly occupied shell transforms as the irrep of the single electron, so the upper open-shell transforms as $e_{2u}$.
The irreps of the lowest-energy excited states are then found by taking the direct product of the irreps of the occupied shells:
\begin{equation}
    \begin{split}
        \Gamma_{\text{Excited State}} &= \Gamma_{\text{core}} \otimes \Gamma_{e_{1g}\text{ shell}} \otimes \Gamma_{e_{2u}\text{ shell}}\\
        & = a_{1g} \otimes e_{1g} \otimes e_{2u} = B_{1u} \oplus B_{2u} \oplus E_{1u}
    \end{split}
    \label{eq:irrepsES}
\end{equation}

Each of these many-body irreps represent both singlet and triplet states. 
Exact diagonalization agrees with these results: at $\alpha=\frac{1}{2}$ and $U=\frac{1}{4}$, the ordering of states is ${^3B_{1u}} < {^3E_{1u}} < {^1B_{2u}} < {^3B_{2u}} < {^1E_{1u}} < {^1B_{1u}}$.

\section{Obstructed Pathway}
\label{sec:ForbiddenPath}

\subsection{Single-Particle Properties}
\label{subsec:ForbiddenPathSP}
We now turn to the obstructed pathway. 
To obtain the single-particle irreps, we block-diagonalize the single-particle Hamiltonian $\bs{h}_{-}(\alpha)$ into the three eigensectors of $C_{3,o}$.
The eigenvalues of $C_{3,o}$ are $\lambda_{k_o}=e^{i k_o}$ with $k_o=\pi,\pm \frac{\pi}{3}$.
The orthogonal transformation into the $C_{3,o}$ basis is given by
\begin{equation}
    \bs{K}_{C_{3,o}}=\frac{1}{\sqrt{3}}\left[
\begin{array}{cccccc}
 0 & 1 & 0 & e^{i\frac{2\pi}{3}} & 0 & e^{-i\frac{2\pi}{3}} \\
 1 & 0 & e^{i\frac{2\pi}{3}} & 0 & e^{-i\frac{2\pi}{3}} & 0 \\
 0 & -1 & 0 & e^{i\frac{\pi}{3}} & 0 & e^{-i\frac{\pi}{3}} \\
 -1 & 0 & e^{i\frac{\pi}{3}} & 0 & e^{-i\frac{\pi}{3}} & 0 \\
 0 & 1 & 0 & 1 & 0 & 1 \\
 1 & 0 & 1 & 0 & 1 & 0 \\
\end{array}
\right],
\end{equation}
and the transformed Hamiltonian is 
\begin{equation}
    \begin{split}
        \bs{h}_{C_{3,o}}(\alpha) & = K_{C_{3,o}}^{\dagger} h_o(\alpha) K_{C_{3,o}} \\
        & =
        \left[
        \begin{array}{ccc}
        \bs{h}_{\pi}(\alpha) & 0 & 0 \\
        0 & \bs{h}_{\frac{\pi}{3}}(\alpha) & 0 \\
        0 & 0 & \bs{h}_{-\frac{\pi}{3}}(\alpha)\\
        \end{array}
        \right],
    \end{split}
\end{equation}
where 
\begin{equation}
    \begin{split}
        \bs{h}_{\pi}(\alpha) &= \left[
        \begin{array}{cc}
         2 \alpha  h & 2 \alpha -1 \\
         2 \alpha -1 & 2 \alpha  h \\
        \end{array}
        \right], \\
        \bs{h}_{\frac{\pi}{3}}(\alpha) &= \left[
        \begin{array}{cc}
         -\alpha h  & \frac{i \sqrt{3} \alpha}{2} + \frac{\alpha}{2}  - 1 \\
        \frac{-i \sqrt{3} \alpha}{2} + \frac{\alpha}{2}  - 1 &  -\alpha h \\
        \end{array}
        \right], \\
        \bs{h}_{-\frac{\pi}{3}}(\alpha)&=\bs{h}_{\frac{\pi}{3}}(\alpha)^*. \\
    \end{split}
\end{equation}

Under $\bs{K}_{C_{3,o}}$, the operators $c_{i,\sigma}^\dagger$ in the site basis transform as
\begin{equation}
    {\pdag{o,i,}{\sigma}} = \sum_{j=1}^{6} [\bs{K}_{C_{3,o}}]_{ij} \cdag{j}{\sigma}
\end{equation}
leading to a new set of operators,
\begin{equation}
    \begin{split}
        \pdag{o,[1,{k_o}],}{\sigma} &= \frac{1}{\sqrt{3}} \sum_{j=1}^{3} e^{i(\pi-j{k_o})} \cdag{2j}{} \\
        \pdag{o,[2,{k_o}],}{\sigma} &= \frac{1}{\sqrt{3}} \sum_{j=1}^{3} e^{i(\pi-j{k_o})} \cdag{2j-1}{}.
    \end{split}
\end{equation}

The block-diagonal, single-particle Hamiltonian $\bs{h}_{C_{3,o}}(\alpha)$ can be diagonalized via a unitary transformation by the matrix of its eigenvectors,
\begin{equation}
    \bs{V}_{o} = \frac{1}{\sqrt{2}}\left[
    \begin{array}{cccccc}
    -1 & 1 & 0 & 0 & 0 & 0 \\
    1 & 1 & 0 & 0 & 0 & 0 \\
    0 & 0 & -\frac{x_o(\alpha)}{y_o(\alpha)} & \frac{x_o(\alpha)}{y_o(\alpha)} & 0 & 0 \\
    0 & 0 & 1 & 1 & 0 & 0 \\
    0 & 0 & 0 & 0 & -\frac{x_o(\alpha)}{z_o(\alpha)} & \frac{x_o(\alpha)}{z_o(\alpha)} \\
    0 & 0 & 0 & 0 & 1 & 1 \\
    \end{array}
    \right],
\end{equation}
where 
\begin{equation}
    \begin{split}
        x_o(\alpha) &= 2 i \sqrt{\alpha ^2-\alpha +1}, \\
        y_o(\alpha) &= -i \alpha -\sqrt{3} \alpha +2 i, \\
        z_o(\alpha) &= -i \alpha +\sqrt{3} \alpha +2 i.
    \end{split}
\end{equation}

This yields the single-particle, MO energies $\epsilon_{1}^{k_o}(\alpha)$, $\epsilon_{2}^{k_o}(\alpha)$
\begin{equation}
    \begin{split}
        \epsilon_{1}^{\pi} (\alpha) &= 1 - 2 \alpha + 2h\alpha \\
        \epsilon_{2}^{\pi} (\alpha) &= - 1 + 2 \alpha + 2h\alpha \\
        \epsilon_{1}^{\frac{-\pi}{3}} (\alpha)= \epsilon_{1}^{\frac{\pi}{3}} (\alpha) &= \sqrt{\alpha ^2-\alpha +1} - h\alpha \\
        \epsilon_{2}^{\frac{-\pi}{3}} (\alpha)= \epsilon_{2}^{\frac{\pi}{3}} (\alpha) &= -\sqrt{\alpha ^2-\alpha +1} -h\alpha
    \end{split}
\end{equation}
and a set of MO-creation operators $\gdag{1,{k_o},}{\sigma}$, $\gdag{2,{k_o},}{\sigma}$ generated by 
\begin{equation}
    \begin{split}
        \gdag{1,{k_o},}{\sigma} &= \frac{1}{\sqrt{2}}(-r_{C_{3,o}}({k_o},\alpha)\pdag{o,[1,{k_o}],}{\sigma} + \pdag{o,[2,k_o],}{\sigma}) \\
        \gdag{2,{k_o},}{\sigma} &= \frac{1}{\sqrt{2}}(r_{C_{3,o}}({k_o},\alpha)\pdag{o,[1,{k_o}],}{\sigma} + \pdag{o,[2,k_o],}{\sigma})
    \end{split}
\end{equation}
where
\begin{equation}
    \begin{split}
        r_{C_{3,o}}(k_o,\alpha)&=C_o(k_o)+(1-C_o(k_o)) \\
        &*\left[\frac{x_o(\alpha)}{y_o(\alpha)}\frac{1+S(k_o)}{2}+\frac{x_o(\alpha)}{z_o(\alpha)}\frac{1-S(k_o)}{2}\right], \\
        C_o(k_o) &= \frac{1-2\cos{k_o}}{3}, \\
        S(k_o) &= \frac{2}{\sqrt{3}}\sin{k_o}.
    \end{split}
\end{equation}

\begin{figure}
    \centering
    \resizebox*{10cm}{!}{\includegraphics{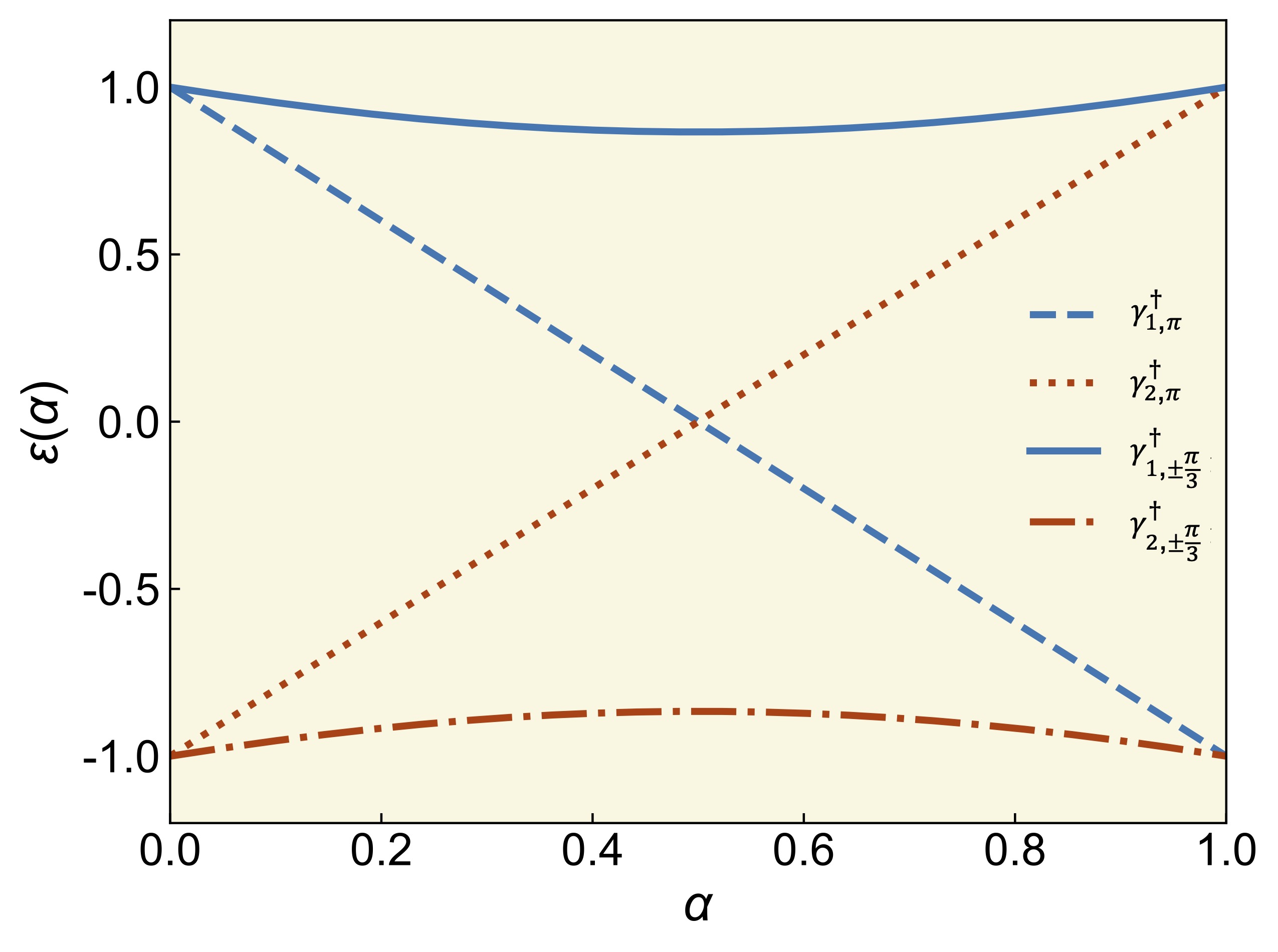}}
    \caption{Single-particle energies of molecular orbitals for the obstructed pathway, where $h=0$. Along the pathway there is a crossing of the frontier orbitals.}
    \label{fig:forbiddenMOenergies}
\end{figure}

Figure \ref{fig:forbiddenMOenergies} displays the single-particle, MO energies of the obstructed pathway as a function of $\alpha$ for $h=0$. 
Unlike the unobstructed pathway, this pathway features three doubly occupied orbitals only when $\alpha \neq \frac{1}{2}$. 
The lowest degenerate MOs $\gdag{2, \pm \frac{\pi}{3}}{}$ remain doubly occupied in the ground state along the entire path. 
However, at $\alpha=\frac{1}{2}$ the otherwise nondegenerate MOs $\gdag{1,\pi}{}$ and $\gdag{2,\pi}{}$ are degenerate, so each orbital is partially occupied, suggesting the presence of multireference correlations.

\subsection{Many-electron states}
To understand the lowest lying many-electron singlet states along the obstructed pathway, it is useful to define approximate eigenstates for $\alpha \rightarrow 0, 1$,
\begin{equation}
\label{eq:forbidden_productstates}
    \begin{split}
        \ket{\Psi_{\alpha < \frac{1}{2}}} &= 
    {\gdag{2,\pi,}{\uparrow}} 
    {\gdag{2,-\frac{\pi}{3},}{\uparrow}} 
    {\gdag{2,\frac{\pi}{3},}{\uparrow}}
    {\gdag{2,\pi,}{\downarrow}}
    {\gdag{2,-\frac{\pi}{3},}{\downarrow}} 
    {\gdag{2,\frac{\pi}{3},}{\downarrow}}  \ket{0} \\
    \ket{\Psi_{\alpha > \frac{1}{2}}} &= 
    {\gdag{1,\pi,}{\uparrow}} 
    {\gdag{2,-\frac{\pi}{3},}{\uparrow}} 
    {\gdag{2,\frac{\pi}{3},}{\uparrow}} 
    {\gdag{1,\pi,}{\downarrow}} 
    {\gdag{2,-\frac{\pi}{3},}{\downarrow}} 
    {\gdag{2,\frac{\pi}{3},}{\downarrow}}  \ket{0}
    \end{split}  
\end{equation}
which correspond to product states where the lowest three MOs are completely filled.

The character of the HOMO switches at $\alpha=\frac{1}{2}$ due to the degeneracy of the frontier orbitals, so we require at least two states.
For any given $\alpha \neq \frac{1}{2}$, one of these product states represents the ground state in the limit of $U \ll \Delta_{H/L}$, where $\Delta_{H/L}$ is the HOMO-LUMO gap, which closes at $\alpha = \frac{1}{2}$ and grows linearly away from the crossing.
From another perspective, $\ket{\Psi_{\alpha > \frac{1}{2}}}$ is related to $\ket{\Psi_{\alpha < \frac{1}{2}}}$ via double excitation for $\alpha < \frac{1}{2}$, while these roles reverse for $\alpha > \frac{1}{2}$.
There exists another low-energy singly excited state for $\alpha \neq \frac{1}{2}$.
We neglect this state since its overlap with the two product states in Eq.~\ref{eq:forbidden_productstates} is negligible for the values of $U$ discussed here.

\begin{figure}
    \centering
    \resizebox*{10cm}{!}{\includegraphics{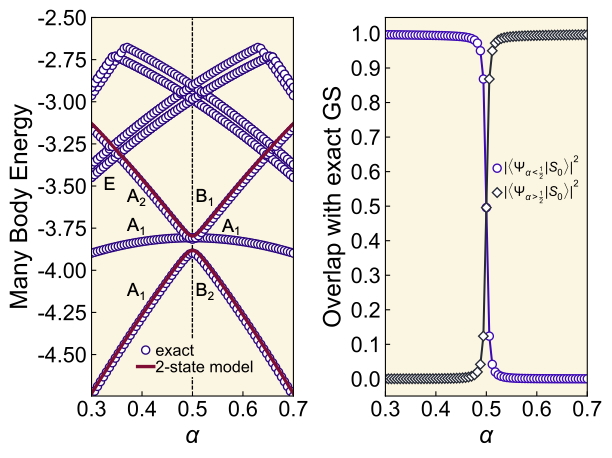}}
    \caption{The left figure shows the singlet many-body energies and two-state model energies with the lowest states irreps. At $\alpha=\frac{1}{2}$ the ordering is $B_2 < A_1 < B_1$. For $0.35 < \alpha < 0.49$ and $0.51 < \alpha < 0.65$, the ordering is $A_1 < A_1 < A_2$. For $\alpha < 0.35$ and $\alpha > 0.65$, the ordering $A_1 < A_1 < E$. The right figure shows the overlaps of product states with the exact ground state.}
    \label{fig:EDforbidden}
\end{figure}

For $\alpha = \frac{1}{2}$ and $U = 0$, the two product states are degenerate with the low-energy singlet, resulting in a threefold degenerate ground state.
For $U > 0$, this degeneracy is lifted and the ground state is unique for all values of $\alpha$ considered here. 
This is illustrated in Figure~\ref{fig:EDforbidden}.

The two product states $\ket{\Psi_{\alpha < \frac{1}{2}}}$ and $\ket{\Psi_{\alpha > \frac{1}{2}}}$ are used to build a two state model, which describes the ground state and a singlet excited state as a function of $\alpha$. 
The Hamiltonian in this two-state basis is given as
\begin{equation}
    \bs{H}_{\textit{eff}} = 
    \left[
    \begin{array}{cc}
    E_1 + 1.5 U + 6\mu & b U \\
    b U & E_2 + 1.5 U + 6\mu \\
    \end{array}
    \right],
\end{equation}
where 
\begin{equation}
    \begin{split}
        E_1 &= -2 + 4\alpha - 4 \sqrt{\alpha^2 - \alpha + 1}, \\
        E_2 &= 2 - 4\alpha - 4 \sqrt{\alpha^2 - \alpha + 1}, 
    \end{split}
\end{equation}
and $b=\frac{1}{6}$.
The energies are given as 
\begin{equation}
    E_\pm = \frac{1}{2}\left( E_1 + E_2 + 3U + 12\mu \pm \sqrt{4b^2U^2 + (E_1 - E_2)^2} \right)
\end{equation}
and the eigenvectors as 
\begin{equation}
    \ket{\pm}= \frac{1}{N_\pm}
    \left[\begin{array}{c}
    f_\pm\\
    1
    \end{array}\right]
\end{equation}
where 
\begin{equation}
    \begin{split}
        f_\pm = \frac{E_1 - E_2 \pm \sqrt{4b^2U^2 + (E_1 -E_2)^2}}{2bU}, \\
        N_\pm = \sqrt{1 + \left| f_\pm \right|^2}.
    \end{split}
\end{equation}
Therefore, the ground state and singlet excited state are 
\begin{equation}
    \begin{split}
        \ket{S_{GS}} &= c_0^- \ket{\Psi_{\alpha < \frac{1}{2}}} + c_1^- \ket{\Psi_{\alpha > \frac{1}{2}}} \\
        \ket{S_{ES}} &= c_0^+ \ket{\Psi_{\alpha < \frac{1}{2}}} + c_1^+ \ket{\Psi_{\alpha > \frac{1}{2}}}
    \end{split}  
\end{equation}
where 
\begin{equation}
    \begin{split}
        c_0^\pm &= \frac{f_\pm}{N_\pm} \\
        c_1^\pm &= \frac{1}{N_\pm}
    \end{split}
\end{equation}

The left panel of Fig.~\ref{fig:EDforbidden} plots the the many-body energies from this two state model and the ten lowest singlet many-body states for $U = \frac{1}{4}$ and $h = 0.05$ obtained via exact diagonalization.
The ground state can always be described by $\ket{S_{GS}}$. 
At $\alpha = \frac{1}{2}$, $\ket{S_{ES}}$ describes the first excited singlet state. 
Near $\alpha = \frac{1}{2}$, $\ket{S_{ES}}$ describes the second excited singlet state.
Further away from $\frac{1}{2}$, $\ket{S_{ES}}$ describes a higher, excited singlet state. 

The overlaps in the right plot of Fig.~\ref{fig:EDforbidden} show that near $\alpha=\frac{1}{2}$ the ground state contains comparable weight from the two product configurations, consistent with a diradicaloid two-configuration description. 
Away from this point, one configuration rapidly becomes dominant.

\subsection{Irreducible Representations}
\label{subsec:ForbiddenIrreps}

We now investigate how this multireference character is represented by the irreps of the many-electron eigenstates.
The symmetries $C_{3,o}$ and $C_{6,o}$ are generators of the double-groups $\tilde{C}_{3v}$ and $\tilde{C}_{6v}$, respectively. 
The symmetry $\sigma_{d,o}$ is also a generator of each of the double-groups, and its matrix representation $D(\sigma_{d,o})$ in the atomic-orbital basis is given in Table~\ref{tab:symmetryMatrices}. 
Unlike the symmetries discussed in Sec.~\ref{sec:SymmetryCriterion}, $D(\sigma_{d,o})$ anti-commutes with TRS, so the eigenvalue of a single-particle state $\ket{n}$ under $\sigma_{d,o}$ is the negative of the complex conjugate of the eigenvalue of its Kramers partner $\ket{\tilde{n}}$: $\lambda_n = -\lambda_{\tilde{n}}^*$.
The character for an occupied set of Kramers partners is then $ \lambda_n \lambda_{\tilde{n}}= - \| \lambda_{\tilde{n}} \| = -1$.
This will result in a nontrivial irrep for a closed-shell ground state of six electrons, which contradicts the argument that a closed-shell ground state will have the trivial irrep. 
The following method avoids the use of $\sigma_{d,o}$ and thus bypasses the nontrivial character problem for a closed-shell system. 

Rather than using the double-groups $\tilde{C}_{3v}$ and $\tilde{C}_{6v}$, we use the single point groups $C_{6v}$ and $C_{12v}$ as covers of these groups, since identifying that $(C_{3,o})^6 = (C_6)^6 = E$ and $(C_{6,o})^{12} = (C_{12})^{12} = E$ allows construction of a one-to-one mapping between the symmetries of the model and the cover groups. 
The generators of each group are $C_n$ and $\sigma_d$, where
\begin{equation}
    D(\sigma_{d})=\left[
\begin{array}{cccccc}
 0 & 0 & 0 & 0 & 0 & 1 \\
 0 & 0 & 0 & 0 & 1 & 0 \\
 0 & 0 & 0 & 1 & 0 & 0 \\
 0 & 0 & 1 & 0 & 0 & 0 \\
 0 & 1 & 0 & 0 & 0 & 0 \\
 1 & 0 & 0 & 0 & 0 & 0 \\
\end{array}
\right],
\end{equation}
which commutes with TRS. 

\begin{table}
    \tbl{Characters and irreducible representations of occupied molecular orbitals for the ground state of the obstructed pathway at $\alpha = \frac{1}{2}$. As discussed in the main text, we identify $C_{12} \equiv C_{6,o}$.}
    {\begin{tabular}{c|ccc|c}
         $C_{12v}$ &  $E$ & $C_{12}$& $\sigma_d$ &  $\Gamma$ \\ \hline
         $\gdag{1,\pi,}{\sigma}$ /  $\gdag{2,\pi,}{\sigma}$&   2&0&  0& $e_3$\\
         $\gdag{2,\pm\frac{\pi}{3},}{\sigma}$&   2&$\sqrt{3}$&  0& $e_1$\\
    \end{tabular}}
    \label{tab:c6vCoverMOs}
\end{table}

Like the unobstructed pathway, we first identify the irreps at the high-symmetry point $\alpha=\frac{1}{2}$ and then then use subduction to identify the irreps at lower symmetries.
The irreps of the occupied MOs under $C_{12v}$ at $\alpha=\frac{1}{2}$ are given in Table~\ref{tab:c6vCoverMOs}.

We use Ford's method to identify the representation of each electron shell~\cite{Ford1972-ug}. 
The core electron shell has irrep $a_1$ and the valence electron shell reduces into a sum of irreps: $a_1 \oplus b_1 \oplus b_2$. 
Consequently, the ground state and the lowest singlet excited states transform as
\begin{equation}
    \begin{split}
        \Gamma_{\text{Ground State}} &= \Gamma_{\text{core}} \otimes \Gamma_{\text{valence}} \\
        &= a_1 \otimes (a_1 \oplus b_1 \oplus b_2) \\
        &= A_1 \oplus B_1 \oplus B_2.
    \end{split}
\end{equation}
Further calculations are necessary to identify the ground state's irrep. 
The approximate ground state $\ket{S_{GS}}$ from the two state model can be used to identify the irrep by operating the point group generators onto the state, revealing that it transforms as $B_2$. The first and second excited singlet states transform as $A_1$ and $B_1$, respectively.

We use subduction to identify the irreps of the lowest many-electron states away from $\alpha=\frac{1}{2}$, using the irreps of the $C_{6v}$ point group. Under subduction to $C_{6v}$, the $B_2$ and $A_1$ states transform as $A_1$, and the $B_1$ state transforms as $A_2$. Thus, at $\alpha=\frac{1}{2}$, the three lowest singlet states have $C_{6v}$ irreps $A_1$, $A_1$, and $A_2$.

We compare the group theoretical predictions with the results from exact diagonalization for $U=\frac{1}{4}$
Within the interval $\alpha=0.5\pm0.01$, the first and second excited singlet states cross. As a result, the ordering of the three lowest singlet irreps becomes $A_1$, $A_2$, and $A_1$ for $0.35<\alpha<0.49$ and $0.51<\alpha<0.65$. Near $\alpha\approx0.35$ and $\alpha\approx0.65$, the state that forms the second excited singlet in these intervals crosses a higher-lying degenerate pair. For $\alpha<0.35$ and $\alpha>0.65$, the second excited singlet transforms as $E$, while the ground and first excited singlet states remain $A_1$ and $A_2$, respectively. 
Away from $\alpha = \frac{1}{2}$, the ground state transforms as the trivial irrep, and therefore there is no symmetry-based obstruction.

\section{Comparison with other multireference descriptors}
\label{sec:RDMdiagnosticHighU}
To compare the symmetry diagnosis with a conventional multireference measure, we evaluate the squared Frobenius norm of the cumulant of the two-particle reduced density matrix, $|{}^2\Delta|^2$, along both pathways for several values of $U$.
This measure is frequently used as a diagnostic of multireference character in related systems~\cite{Ganoe2024-rm, Juhasz2006-hd}.

\begin{figure}
    \centering
    \resizebox{10cm}{!}{\includegraphics{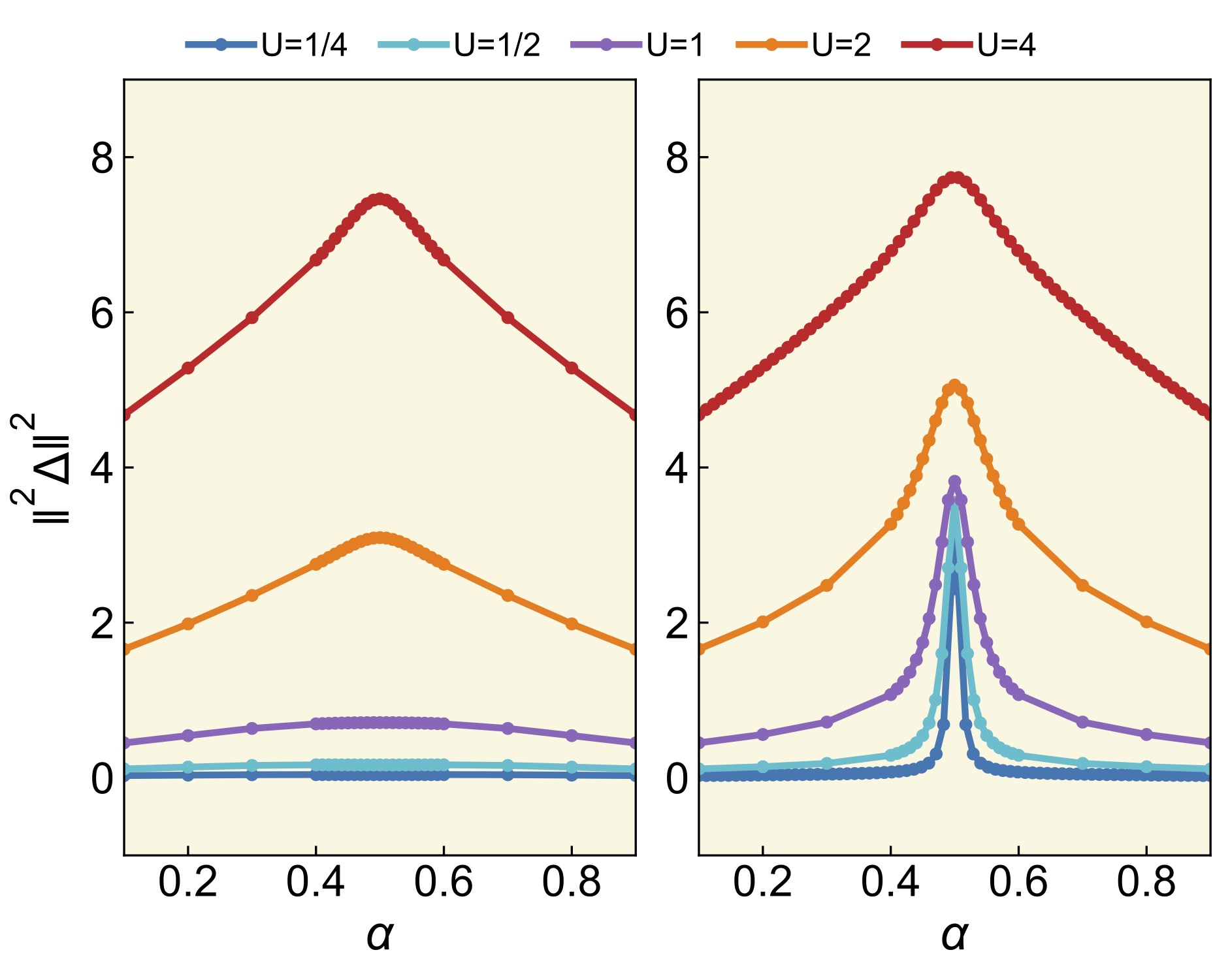}}
    \caption{The squared Frobenius norm of the cumulant of the 2-RDM ($\| {}^2 \Delta\|^2$) for different values of $U$ along the unobstructed (left) and obstructed (right) pathways.}
    \label{fig:RDMdiagnostics}
\end{figure}

Figure~\ref{fig:RDMdiagnostics} shows $\| {}^2 \Delta\|^2$ along each pathway for different values of $U$. 
For $U \leq 1$ along the unobstructed pathway, the norm of the cumulant is less than $1$ with a maximum value of $\| {}^2 \Delta\|^2 \simeq 0.711$ at $\alpha = \frac{1}{2}$ and $U = 1$.
This indicates that the ground state at these points can be qualitatively described by a product state.
The obstructed pathway acts similarly for these values of $U$ except within the region of $\alpha = 0.5 \pm 0.1$ where $\| {}^2 \Delta\|^2$ approaches $3.82$ for $U = 1$, indicating significant multireference character. 
In this regime, the density-matrix diagnostic identifies the same distinction seen in the irrep analysis and in the exact diagonalization overlaps, i.e., the unobstructed pathway remains qualitatively closed-shell, whereas the obstructed pathway develops pronounced two-configuration character at the high-symmetry point.

For $U = 2$, both pathways show significant multireference character for all $\alpha$, as the cumulant peaks at $\alpha=\frac{1}{2}$ and takes on a minimal value  $\| {}^2 \Delta\|^2 \simeq 1.65$.
However, $\| {}^2 \Delta\|^2$ at $\alpha = \frac{1}{2}$ on the unobstructed pathway remains smaller than its corresponding point on the obstructed pathway for $U \leq 2$. 
Moreover, for $U=4$, $\| {}^2 \Delta\|^2 > 4.67$ for all values of $\alpha$ along both pathways, resulting in a loss of distinction between these pathways in terms of this descriptor.

From the group theoretical point of view, the irreps for the single-particle and many-body states are not dependent on the value of $U$~\cite{Lieb1989-xw,Muechler-Mobius}. 
This reveals a limitation of using irreps as a multireference diagnostic. 
For example, $\| {}^2 \Delta\|^2$ has shown that the ground state of the unobstructed pathway for $U=4$ is multireference, but the trivial irreps of the state would suggest otherwise. 
Therefore, the group theoretical approach is limited for $U \gg \Delta_{\alpha}$. 

An additional consideration is the fact that our group theoretical approach only included fully and partially occupied orbitals, neglecting the unoccupied orbitals at higher energies. 
This approach is well justified if the gap between the selected sets of orbitals and those not considered is much larger than $U$. 
For larger values of $U$, excitations into these high-lying orbitals become significant. 
Inclusion of these orbitals into the group theoretical setting is possible without adjustments, but they are generally less significant as the number of possible electronic states grows exponentially with the number of orbitals involved. 
For reactions involving only two active orbitals, as in the diradical or stationary states of symmetry-forbidden reactions or anti-aromatic molecules such as cyclobutadiene, we believe the group theoretical approach is well justified.

For example, the isomerization of cyclobutadiene features a transition state at the $D_{4h}$ symmetric geometry that transforms as the $B_{1g}$ irrep of $D_{4h}$, while it transforms as the $A_g$ irrep of $D_{2h}$ at the other points along the pathway. 
Close to the $D_{4h}$ symmetric point, multireference correlations become significant, requiring the use of multireference approaches~\cite{Monino2022-ls}.

\section{Discussion}
\label{sec:Discussion}
We have shown that the irreps of many-electron states can diagnose a symmetry obstruction to a time-reversal-preserving closed-shell description in a six-site Hubbard model with two pathways. An unobstructed pathway is characterized by a finite HOMO-LUMO gap and a trivial ground-state irrep, whereas the obstructed pathway possesses a HOMO-LUMO degeneracy in the form of a linear crossing. For small interaction strengths, we find a two-configuration singlet ground state, and a nontrivial many-electron irrep. In the weak-to-intermediate interaction regime, this symmetry-based diagnosis is consistent with exact diagonalization, a two-state effective model, and the two-particle cumulant.

From a group-theoretical point of view, the single-particle and many-electron irreps in our model are independent of $U$~\cite{Lieb1989-xw,Muechler-Mobius}. This exposes an important limitation of irreps as a multireference diagnostic. For example, the multireference metric $|{}^2\Delta|^2$ shows that the ground state along the unobstructed pathway at $U=4$ has multireference character, whereas its trivial irrep does not diagnose this. Thus, the irrep-based criterion is most useful when multireference character arises from symmetry-enforced near degeneracies, and it becomes less informative when correlations are driven by large $U$.

A second limitation is that our group-theoretical analysis includes only the fully and partially occupied orbitals, while neglecting higher-energy unoccupied orbitals. This approximation is justified when the gap between the selected orbitals and the neglected orbitals is much larger than $U$. For larger $U$, excitations into these higher-lying orbitals can become important. These orbitals could be included in the same group-theoretical framework without formal modification, but doing so is generally less practical because the number of possible electronic states grows exponentially with the number of active orbitals. For reactions dominated by two active orbitals, such as diradicals, stationary points of symmetry-forbidden reactions, or antiaromatic molecules such as cyclobutadiene, the group-theoretical approach remains well justified and practical for screening purposes.

Most chemically relevant molecules are not of high-symmetry. However, the analysis of high-symmetry models remains useful for identifying which symmetries obstruct a single-reference description in practice. Models with artificially high symmetry make this especially transparent, because individual symmetry breaking can be introduced in a controlled way. In the present model, breaking $C_{6,o}$ symmetry does not by itself remove the obstruction at $\alpha=\frac{1}{2}$ for the obstructed pathway, provided the relevant mirror symmetry $\sigma_v$ is preserved. The ground state remains distinguished from the trivial irrep by its $\sigma_v$ and $C_{12}$ eigenvalues. By contrast, sufficiently strong breaking of $\sigma_v$ is expected to remove the obstruction, because the crossing between the two MOs is protected by $\sigma_v$ symmetry. However, upon breaking of $\sigma_v$, the degeneracy can be lifted, and the symmetry-based distinction between the correlated ground state and a trivial single-reference state is no longer enforced.

The broader significance of the connection between irreps and multireference correlation is that irreps also determine spectroscopic selection rules. High-symmetry diradicaloid and antiaromatic states therefore provide natural settings in which nontrivial many-electron irreps may lead to spectroscopic signatures distinct from those of symmetry-trivial closed-shell states.


\section*{Disclosure statement}
The authors report there are no competing interests to declare.

\section*{Funding}

This work was supported by the U.S. Department of Energy, Office of Science, Office of Basic Energy Sciences, CPIMS program, under Award DE-SC0025352.




\bibliographystyle{tfo}
\bibliography{molphys_references}

\appendix

\section{Formalism: Calculating Many-Electron Irreducible Representations}

As an example of how to calculate many-electron irreps for an open-shell system, we consider a single-excitation on the unobstructed pathway at $\alpha = \frac{1}{2}$. 
This single-excitation results in three electrons in $e_{1g}$ and one electron in $e_{2u}$.
For the lower open-shell, we could naively take the direct product of the irreps of the electrons and use the reduction formula
\begin{equation}
    N_\alpha = \frac{1}{h} \sum_G n_G \cdot \chi_\alpha^\ast(G) \cdot \chi_R(G)
\end{equation}
to find the sum of possible irreps: 
\begin{equation}
    \begin{split}
    \Gamma_{e_{1g} \text{shell}} & = e_{1g} \otimes e_{1g} \otimes e_{1g} \\
    & = B_{1g} \oplus B_{2g} \oplus 3E_{1g}.
    \end{split}
\end{equation}
However, not all of these irreps are appropriate fermionic states for three electrons in doubly degenerate orbitals, as we will explain below.

For two electrons occupying a set of MOs with the same irrep, taking the direct product of an irrep by itself results in a sum of the symmetrized product, antisymmetrized product, and mixed-symmetric product.
Since a singlet state has a spin antisymmetric wavefunction, the irrep of interest is the symmetric product, which represents the spatial part of the wavefunction.
An antisymmetrized product of the spatial part of the wavefunction corresponds to a triplet, since the spin part is symmetric. 
The mixed-symmetric product is irrelevant for two electrons since they cannot form a mixed-symmetric orbital or spin state. 
When finding the product of more than two electrons, the spatial part of the wavefunction may have mixed-symmetry since the spin part may have mixed-symmetry, e.g., for a doublet.

For these excited states, the three-electron shell has mixed-symmetric character for both the spatial and spin parts of the wavefunction so that the total wavefunction transforms with antisymmetric character. 
The one electron shell must also have antisymmetric character. 
Together, the two shells can form singlet or triplet states, resulting in 16 possible electronic configurations.
To identify the correct number of singlet and triplet states and their irreps, we use Ford's method, which employs the symmetric group and the language of Young tableaux~\cite{Ford1972-ug, Schensted1967-tk}.

An antisymmetric wavefunction with $n$-electrons must transform as the $\{ 1^n \}$ irrep of the symmetric group $S_n$.
Empty Young tableaux, called Young diagrams, are a visual way of representing the irreps of $S_n$, such that a column Young diagram is the antisymmetic irrep, a row diagram is a symmetric irrep, and any other diagram is a mixed-symmetric irrep. 
Once the Young diagram for the spatial part of the wavefunction is obtained for $n$-electrons, the appropriate equation for finding the product of $n$ point group irreps is given by the character table for the symmetric group $S_n$. 
We demonstrate this method for the three-electron $e_{1g}$ shell.

The only allowed Young tableaux for three spins in two orbitals is 
\begin{equation}
    \ytableaushort{\uparrow \uparrow,\downarrow}
    \label{eq:ytab3spins}
\end{equation}
The spatial part of the wavefunction will have a Young diagram conjugate to that for the spins since they must have opposite symmetry.
In this case, the conjugate diagram has the same shape. 
In the absence of spin-orbit coupling, we can ignore the step for making the Young tableaux for the spins and begin at this spatial Young diagram step~\cite{Matsen1970-ch}.
\begin{equation}
    \ydiagram{2,1}
    \label{eq:ytab3orbitals}
\end{equation}
The diagram in Eq.~\ref{eq:ytab3orbitals} represents the $\{2, 1\}$ irrep, which is the mixed-symmetric irrep in $S_3$. 
Following Ford's method and using the $S_3$ character table, we obtain Eq.~\ref{eq:charS3} from the $\{2, 1\}$ irrep and use it to find the correct mixed-symmetric product of three of the same irreps:
\begin{equation}
    \chi(G: \text{doublet}) = \frac{1}{3} \left\{ \chi(G)^3 - \chi(G^3)\right\}
    \label{eq:charS3}
\end{equation}
Here, $\chi(G:\text{doublet})$ is the character under a symmetry $G$ for a doublet state, $\chi(G)$ is the character of the point group irrep under $G$, and $\chi(G^3)$ is the character of $G^3$.

Since the single-particle states transform as $e_{1g}$, which has gerade symmetry, the irrep of the shell will also have gerade symmetry. 
We can therefore ignore the inversion and improper rotation symmetries and simplify our calculations such that only the $D_6$ subgroup of $D_{6h}$ is considered. 
We utilize the $D_6$ character table and follow Ford's method outlined in Ref.~\cite{Ford1972-ug}, the calculations for which are organized in Table~\ref{tab:d6_character_table}, which reveal that the three-electron shell transforms as $e_{1g}$.
A singly occupied shell transforms as the irrep of the single electron, so the upper open-shell transforms as $e_{2u}$.
The irreps of the lowest-energy excited states are now found by taking the direct product of the irreps of the occupied shells:
\begin{equation}
    \begin{split}
        \Gamma_{\text{Excited State}} &= \Gamma_{\text{core}} \otimes \Gamma_{e_{1g}\text{ shell}} \otimes \Gamma_{e_{2u}\text{ shell}}\\
        & = a_{1g} \otimes e_{1g} \otimes e_{2u} = B_{1u} \oplus B_{2u} \oplus E_{1u}
    \end{split}
    \label{eq:irrepsES}
\end{equation}

The resulting sum in Eq.~\ref{eq:irrepsES} contains all possible singlet and triplet excited states derived from a single electronic excitation to the LUMO from the HOMO~\cite{Ellis1971-pk}.
This is due to the fact that the product of two of the same representation can result in a sum of irreps, upon which the singlet or triplet states must be identified, and for such was the benefit of using the Ford method. 
However, when the product of two different representations is taken, the resulting representations can be assigned to both singlet and triplet states since the spin of the electrons, which occupy different orbital shells as implied by the different representations, is not restricted by the Pauli exclusion principle. 
In the cases where there is a sum of irreps, the state of interest and the next lowest energy states are each assigned to one of the representations. 
For this single excitation, there are four low-lying singlet states and twelve low-lying triplet states with these irreps, totaling the expected 16 possible singly-excited states.

\begin{table}
    \tbl{\textbf{Example for organization of calculations using Ford's method.} The table organizes the calculations for finding the irrep of the excited state's lower open-shell at $\alpha=\frac{1}{2}$ on the unobstructed pathway. The point group used is $\bs{D}_6$. $G$ represents a symmetry element and $G^3$ is the element cubed. For example, $(C_6)^3 \equiv C_2$. The irrep of an electron in this shell is $e_1$, and its characters are given in the next row of the table. $\chi(G)$ is the character of the symmetry $G$ under the irrep of interest, $e_1$, and $\chi(G^3)$ is the character of the symmetry $G^3$. For example, in the column for the element $C_6$, $\chi(G) = \chi(C_6)$ in $e_1$, which is $+1$. Similarly, $\chi(G^3) = \chi(C_2)$ which is $-2$ under $e_1$. Once the characters are tabulated for $G$ and $G^3$ under $e_1$, the character of three electrons with $e_1$ symmetry and which form a doublet is calculated for each symmetry element using Eq.~\ref{eq:charS3}. The resulting characters inform the reducible representation, which can be reduced into a sum of irreps. In this case, the irrep of this partially filled electron shell is $e_1$.}
    {\begin{tabular}{rccccccc}\toprule
        \textbf{D$_6$} & \textbf{E} & \textbf{2C$_6$(z)} & \textbf{2C$_3$(z)} & \textbf{C$_2$(z)} & \textbf{3C'$_2$} & \textbf{3C''$_2$}  & \\ \midrule
        $G$ & E & C$_6$ & C$_3$ & C$_2$ & $C^{'}_2$& $C^{''}_2$& \\
        $G^3$ & E & C$_2$ & E & C$_2$ & $C^{'}_2$& $C^{''}_2$& \\ \midrule
        \multicolumn{7}{c}{} & \\ \midrule
  $e_1$& +2 & +1 & -1 & -2 & 0 &0  & \\ \midrule
        $\chi(G)$ in $e_{1}$& +2 & +1 & -1 & -2 & 0 & 0  & \\
        $\chi(G^3)$ in $e_{1}$& +2 & -2 & +2 & -2 & 0 & 0  & \\ \midrule
        $\chi(G: \text{doublet})$& +2& +1 & -1& -2& 0& 0& $e_1$\\ \bottomrule
    \end{tabular}}
\label{tab:d6_character_table}
\end{table}

\section{$\tilde{C}_{3v}$ and $\tilde{C}_{6v}$ Double Groups: Character and Multiplication Tables}

\begin{table}
    \setlength{\tabcolsep}{4pt} 
    \tbl{\textbf{Character table for} $\bs{\tilde{C}_{3v}}$. $\tilde{C}_{3v}$ is the double group of $C_{3v}$ with twice the number of symmetry elements. $A_1$, $A_2$, and $E$ are vector representations of both the single group and double group. $E_{1/2}$, ${}^1E_{3/2}$, and ${}^2E_{3/2}$ are spinor representations, which are only in the double group. The symmetry elements $\tilde{E}$, $2\tilde{C}_3$, and $3\tilde{\sigma}_v$ are unique to the double group. Each of the symmetry elements with a tilde, $\tilde{C}$, are related to their corresponding elements, $C$, by $-E$, i.e., their matrix representations are the negative of the matrix representations of their corresponding symmetries. This character table was adapted from the corresponding table given by Altmann in \textit{Point-Group Theory Tables} according to the instructions given by the authors~\cite{Altmann1994-ti}.}
    {\begin{tabular}{l @{\hspace{12pt}} rrr @{\hspace{10pt}} rrr}
        \toprule \addlinespace[4pt]
        $\bs{\tilde{C}_{3v}}$ & $E$ & $2C_3$ & $3\sigma_v$ & $\tilde{E}$ & $2\tilde{C}_3$ & $3\tilde{\sigma}_v$ \\ 
        \midrule \addlinespace[4pt]
        $A_1$ & $1$ &  $1$& $1$ & $1$ & $1$ & $1$  \\ \addlinespace[1pt]
        $A_2$ & $1$ & $1$ &  $-1$ &  $1$& $1$ & $-1$  \\ \addlinespace[1pt]
        $E$ &  $2$ &  $-1$& $0$ &  $2$& $-1$ & $0$ \\ \addlinespace[1pt]
        $E_{1/2}$ & $2$ & $1$ & $0$ & $-2$ & $-1$ & $0$ \\ \addlinespace[1pt]
        $^1E_{3/2}$ & $1$ & $-1$ & $i$ & $-1$ & $1$ & $-i$ \\ \addlinespace[1pt]
        $^2E_{3/2}$ & $1$ & $-1$ & $-i$ & $-1$ & $1$ & $i$ \\ \addlinespace[1pt]
        \bottomrule 
    \end{tabular}}
    \label{tab:c3vDGcharacter}
\end{table}

\begin{table}
    \setlength{\tabcolsep}{4pt} 
    \tbl{\textbf{Multiplication table for} $\bs{\tilde{C}_{3v}}$. This multiplication table for the double-group was constructed using the multiplication and factor tables of $C_{3v}$ according to the instructions given by Altmann in \textit{Point-Group Theory Tables}~\cite{Altmann1994-ti}. The product of any two symmetry elements $g_i g_j$ is given by $g_k$ at the intersection of the row $g_i$ and the column $g_j$. The top left block, the ``basic block,'' is the product of the multiplication and factor tables. It is identical to the bottom right block. The other two blocks, which are identical, are the ``complement'' of the basic block, i.e., it is the basic block multiplied by $\tilde{E}$. The ordering of the elements in the header row and header column is according to the classes of $\tilde{C}_{3v}$, which are given in the header of the character table of $\tilde{C}_{3v}$.}
    {\begin{tabular}{l @{\hspace{16pt}} cccccc @{\hspace{12pt}} cccccc}
        \toprule \addlinespace[4pt]
        $\bs{\tilde{C}_{3v}}$ & $\bs{E}$ & $\bs{C_3^+}$ & $\bs{C_3^-}$ & $\bs{\sigma_{d1}}$ & $\bs{\sigma_{d2}}$ & $\bs{\sigma_{d3}}$ & $\bs{\tilde{E}}$ & $\bs{\tilde{C}_3^+}$ & $\bs{\tilde{C}_3^-}$ & $\bs{\tilde{\sigma}_{d1}}$ & $\bs{\tilde{\sigma}_{d2}}$ & $\bs{\tilde{\sigma}_{d3}}$ \\
        \midrule \addlinespace[6pt]
        $\bs{E}$ & $E$ & $C_3^+$ & $C_3^-$ & $\sigma_{d1}$ & $\sigma_{d3}$ & $\sigma_{d3}$ & $\tilde{E}$ & $\tilde{C}_3^+$ & $\tilde{C}_3^-$ & $\tilde{\sigma}_{d1}$ & $\tilde{\sigma}_{d2}$ & $\tilde{\sigma}_{d3}$ \\ \addlinespace[2pt]
        
        $\bs{C_3^+}$ & $C_3^+$ & $\tilde{C}_3^-$ & $E$ & $\tilde{\sigma}_{d3}$ & $\tilde{\sigma}_{d1}$ & $\tilde{\sigma}_{d2}$ & $\tilde{C}_3^+$ & ${C}_3^-$ & $\tilde{E}$ & ${\sigma}_{d3}$ & ${\sigma}_{d1}$ & ${\sigma}_{d2}$ \\ \addlinespace[2pt]
        
        $\bs{C_3^-}$ & $C_3^-$ & $E$ & $\tilde{C}_3^+$ & $\tilde{\sigma}_{d2}$ & $\tilde{\sigma}_{d3}$ & $\tilde{\sigma}_{d1}$ & $\tilde{C}_3^-$ & $\tilde{E}$ & ${C}_3^+$ & ${\sigma}_{d2}$ & ${\sigma}_{d3}$ & ${\sigma}_{d1}$ \\ \addlinespace[2pt]
        
        $\bs{\sigma_{d1}}$ & $\sigma_{d1}$ & $\tilde{\sigma}_{d2}$ & $\tilde{\sigma}_{d3}$ & $\tilde{E}$ & $C_3^+$ & $C_3^-$ & $\tilde{\sigma}_{d1}$ & ${\sigma}_{d2}$ & ${\sigma}_{d3}$ & ${E}$ & $\tilde{C}_3^+$ & $\tilde{C}_3^-$ \\ \addlinespace[2pt]
        
        $\bs{\sigma_{d2}}$ & $\sigma_{d2}$ & $\tilde{\sigma}_{d3}$ & $\tilde{\sigma}_{d1}$ & $C_3^-$ & $\tilde{E}$ & $C_3^+$ & $\tilde{\sigma}_{d2}$ & ${\sigma}_{d3}$ & ${\sigma}_{d1}$ & $\tilde{C}_3^-$ & ${E}$ & $\tilde{C}_3^+$  \\ \addlinespace[2pt]
        
        $\bs{\sigma_{d3}}$ & $\sigma_{d3}$ & $\tilde{\sigma}_{d1}$ & $\tilde{\sigma}_{d2}$ & $C_3^+$ & $C_3^-$ & $\tilde{E}$ & $\tilde{\sigma}_{d3}$ & ${\sigma}_{d1}$ & ${\sigma}_{d2}$ & $\tilde{C}_3^+$ & $\tilde{C}_3^-$ & ${E}$\\ \addlinespace[10pt]
        
        $\bs{\tilde{E}}$ & $\tilde{E}$ & $\tilde{C}_3^+$ & $\tilde{C}_3^-$ & $\tilde{\sigma}_{d1}$ & $\tilde{\sigma}_{d2}$ & $\tilde{\sigma}_{d3}$ & $E$ & $C_3^+$ & $C_3^-$ & $\sigma_{d1}$ & $\sigma_{d3}$ & $\sigma_{d3}$ \\ \addlinespace[2pt]
        
        $\bs{\tilde{C}_3^+}$ & $\tilde{C}_3^+$ & ${C}_3^-$ & $\tilde{E}$ & ${\sigma}_{d3}$ & ${\sigma}_{d1}$ & ${\sigma}_{d2}$ & $C_3^+$ & $\tilde{C}_3^-$ & $E$ & $\tilde{\sigma}_{d3}$ & $\tilde{\sigma}_{d1}$ & $\tilde{\sigma}_{d2}$ \\ \addlinespace[2pt]
        
        $\bs{\tilde{C}_3^-}$ & $\tilde{C}_3^-$ & $\tilde{E}$ & ${C}_3^+$ & ${\sigma}_{d2}$ & ${\sigma}_{d3}$ & ${\sigma}_{d1}$  & $C_3^-$ & $E$ & $\tilde{C}_3^+$ & $\tilde{\sigma}_{d2}$ & $\tilde{\sigma}_{d3}$ & $\tilde{\sigma}_{d1}$  \\ \addlinespace[2pt]
        
        $\bs{\tilde{\sigma}_{d1}}$ &  $\tilde{\sigma}_{d1}$ & ${\sigma}_{d2}$ & ${\sigma}_{d3}$ & ${E}$ & $\tilde{C}_3^+$ & $\tilde{C}_3^-$  & $\sigma_{d1}$ & $\tilde{\sigma}_{d2}$ & $\tilde{\sigma}_{d3}$ & $\tilde{E}$ & $C_3^+$ & $C_3^-$ \\ \addlinespace[2pt]
        
        $\bs{\tilde{\sigma}_{d2}}$ & $\tilde{\sigma}_{d2}$ & ${\sigma}_{d3}$ & ${\sigma}_{d1}$ & $\tilde{C}_3^-$ & ${E}$ & $\tilde{C}_3^+$ & $\sigma_{d2}$ & $\tilde{\sigma}_{d3}$ & $\tilde{\sigma}_{d1}$ & $C_3^-$ & $\tilde{E}$ & $C_3^+$ \\ \addlinespace[2pt]

        $\bs{\tilde{\sigma}_{d3}}$ & $\tilde{\sigma}_{d3}$ & ${\sigma}_{d1}$ & ${\sigma}_{d2}$ & $\tilde{C}_3^+$ & $\tilde{C}_3^-$ & ${E}$ & $\sigma_{d3}$ & $\tilde{\sigma}_{d1}$ & $\tilde{\sigma}_{d2}$ & $C_3^+$ & $C_3^-$ & $\tilde{E}$ \\ \addlinespace[4pt]
        \bottomrule
    \end{tabular}}
    \label{tab:C3vDGmultiplication}
\end{table}

\begin{table}
    \setlength{\tabcolsep}{4pt} 
    \tbl{\textbf{Character table for} $\bs{\tilde{C}_{6v}}$. $\tilde{C}_{6v}$ is the double-group of $C_{6v}$ with twice the number of symmetry elements. $A_1$, $A_2$, $B_1$, $B_2$, $E_1$, and $E_2$ are vector representations of both the single- and double-groups. $E_{1/2}$, $E_{3/2}$, and $E_{5/2}$ are spinor representations, which are only in the double-group. The symmetry elements $\tilde{E}$, $2\tilde{C}_6$, $2\tilde{C}_3$, $\tilde{C}_2$, $3\tilde{\sigma}_d$, and $3\tilde{\sigma}_v$ are unique to the double-group. Each of the symmetry elements with a tilde, $\tilde{C}$, are related to their corresponding elements, $C$, by $-E$, i.e., their matrix representations are the negative of the matrix representations of their corresponding symmetries. Element which are put together in parentheses belong to the same class. Unlike $\tilde{C}_{3v}$, this group has some classes which contain both single- and double-group elements. Therefore, $\tilde{C}_{6v}$ does not have twice the number of classes as $C_{6v}$. This character table was adapted from the corresponding table given by Altmann in \textit{Point-Group Theory Tables} according to the instructions given by the authors~\cite{Altmann1994-ti}.}
    {\begin{tabular}{l @{\hspace{12pt}} rrrrrrrrr}
        \toprule \addlinespace[4pt]
        \(\bs{\tilde{C}_{6v}}\) 
            & \(E\) 
            & \(\tilde{E}\) 
            & \(2C_6\) 
            & \(2\tilde{C}_6\) 
            & \(2C_3\) 
            & \(2\tilde{C}_3\) 
            & \((C_2,\tilde{C}_2)\) 
            & \((3\sigma_d,3\tilde{\sigma}_d)\) 
            & \((3\sigma_v,3\tilde{\sigma}_v)\) \\
        \midrule \addlinespace[4pt]
        $A_1$ & $1$ & $1$ & $1$ & $1$ & $1$ & $1$ & $1$ & $1$ & $1$ \\ \addlinespace[1pt]
        $A_2$ & $1$ & $1$ & $1$ & $1$ & $1$ & $1$ & $1$ & $-1$ & $-1$ \\ \addlinespace[1pt]
        $B_1$ & $1$ & $1$ & $-1$ & $-1$ & $1$ & $1$ & $-1$ & $-1$ & $1$ \\ \addlinespace[1pt]
        $B_2$ & $1$ & $1$ & $-1$ & $-1$ & $1$ & $1$ & $-1$ & $1$ & $-1$ \\ \addlinespace[1pt]
        $E_1$ & $2$ & $2$ & $1$ & $1$ & $-1$ & $-1$ & $-2$ & $0$ & $0$ \\ \addlinespace[1pt]
        $E_2$ & $2$ & $2$ & $-1$ & $-1$ & $-1$ & $-1$ & $2$ & $0$ & $0$ \\ \addlinespace[1pt]
        $E_{1/2}$ & $2$ & $-2$ & $\sqrt{3}$ & $-\sqrt{3}$ & $1$ & $-1$ & $0$ & $0$ & $0$ \\ \addlinespace[1pt]
        $E_{3/2}$ & $2$ & $-2$ & $0$ & $0$ & $-2$ & $2$ & $0$ & $0$ & $0$ \\ \addlinespace[1pt]
        $E_{5/2}$ & $2$ & $-2$ & $-\sqrt{3}$ & $\sqrt{3}$ & $1$ & $-1$ & $0$ & $0$ & $0$ \\ \addlinespace[1pt]
        \bottomrule
    \end{tabular}}
    \label{tab:c6vDGcharacter}
\end{table}

\setcounter{table}{3}
\begin{sidewaystable}
\setlength{\tabcolsep}{4pt}
\tbl{\textbf{Multiplication table for} $\bs{\tilde{C}_{6v}}$. This multiplication table for the double-group was constructed using the multiplication and factor tables of $C_{6v}$ according to the instructions given by Altmann in \textit{Point-Group Theory Tables}~\cite{Altmann1994-ti}. The product of any two symmetry elements $g_i g_j$ is given by $g_k$ at the intersection of the row $g_i$ and the column $g_j$. The top left block, the ``basic block,'' is the product of the multiplication and factor tables. It is identical to the bottom right block. The other two blocks, which are identical, are the ``complement'' of the basic block, i.e., it is the basic block multiplied by $\tilde{E}$. The ordering of the elements in the header row and header column is according to the classes of $C_{6v}$ and then their corresponding double-group symmetry elements.}
{\begin{tabular}{l @{\hspace{16pt}} cccccccccccc @{\hspace{12pt}}  cccccccccccc}
.\toprule \addlinespace[4pt]
        $\bs{\tilde{C}_{6v}}$ & $\bs{E}$ & $\bs{C_6^+}$ & $\bs{C_6^-}$ & $\bs{C_3^+}$ & $\bs{C_3^-}$ & $\bs{C_2}$ & $\bs{\sigma_{d1}}$ & $\bs{\sigma_{d2}}$ & $\bs{\sigma_{d3}}$ & $\bs{\sigma_{v1}}$ & $\bs{\sigma_{v2}}$ & $\bs{\sigma_{v3}}$ & $\bs{\tilde{E}}$ & $\bs{\tilde{C}_6^+}$ & $\bs{\tilde{C}_6^-}$ & $\bs{\tilde{C}_3^+}$ & $\bs{\tilde{C}_3^-}$ & $\bs{\tilde{C}_2}$ & $\bs{\tilde{\sigma}_{d1}}$ & $\bs{\tilde{\sigma}_{d2}}$ & $\bs{\tilde{\sigma}_{d3}}$ & $\bs{\tilde{\sigma}_{v1}}$ & $\bs{\tilde{\sigma}_{v2}}$ & $\bs{\tilde{\sigma}_{v3}}$ \\
        \midrule \addlinespace[6pt]
        $\bs{E}$  & $E$ & $C_6^+$ & $C_6^-$ & $C_3^+$ & $C_3^-$ & $C_2$ & $\sigma_{d1}$ & $\sigma_{d2}$ & $\sigma_{d3}$ & $\sigma_{v1}$ & $\sigma_{v2}$ & $\sigma_{v3}$ & $\tilde{E}$ & $\tilde{C}_6^+$ & $\tilde{C}_6^-$ & $\tilde{C}_3^+$ & $\tilde{C}_3^-$ & $\tilde{C}_2$ & $\tilde{\sigma}_{d1}$ & $\tilde{\sigma}_{d2}$ & $\tilde{\sigma}_{d3}$ & $\tilde{\sigma}_{v1}$ & $\tilde{\sigma}_{v2}$ & $\tilde{\sigma}_{v3}$\\ \addlinespace[2pt]
        
        $\bs{C_6^+}$  & $C_6^+$ & $C_3^+$ & $E$ & $C_2$ & $C_6^-$ & $\tilde{C}_3^-$ & $\tilde{\sigma}_{v2}$ & $\tilde{\sigma}_{v3}$ & $\tilde{\sigma}_{v1}$ & $\sigma_{d2}$ & $\sigma_{d3}$ & $\sigma_{d1}$ & $\tilde{C}_6^+$ & $\tilde{C}_3^+$ & $\tilde{E}$  & $\tilde{C}_2$ & $\tilde{C}_6^-$ & $C_3^-$ & $\sigma_{v2}$ & $\sigma_{v3}$ & $\sigma_{v1}$ & $\tilde{\sigma}_{d2}$ & $\tilde{\sigma}_{d3}$ & $\tilde{\sigma}_{d1}$ \\ \addlinespace[2pt]
        
        $\bs{C_6^-}$  & $C_6^-$ & $E$ & $C_3^-$ & $C_6^+$ & $\tilde{C}_2$ & $C_3^+$ & $\sigma_{v3}$ & $\sigma_{v1}$ & $\sigma_{v2}$ & $\tilde{\sigma}_{d3}$ & $\tilde{\sigma}_{d1}$ & $\tilde{\sigma}_{d2}$ & $\tilde{C}_6^-$ & $\tilde{E}$ & $\tilde{C}_3^-$ & $\tilde{C}_6^+$ & $C_2$ & $\tilde{C}_3^+$ & $\tilde{\sigma}_{v3}$ & $\tilde{\sigma}_{v1}$ & $\tilde{\sigma}_{v2}$ & $\sigma_{d3}$ & $\sigma_{d1}$ & $\sigma_{d2}$ \\ \addlinespace[2pt]
        
        $\bs{C_3^+}$  & $C_3^+$ & $C_2$ & $C_6^+$ & $\tilde{C}_3^-$ & $E$ & $\tilde{C}_6^-$ & $\tilde{\sigma}_{d3}$ & $\tilde{\sigma}_{d1}$ & $\tilde{\sigma}_{d2}$ & $\tilde{\sigma}_{v3}$ & $\tilde{\sigma}_{v1}$ & $\tilde{\sigma}_{v2}$ & $\tilde{C}_3^+$ & $\tilde{C}_2$ & $\tilde{C}_6^+$ & $C_3^-$ & $\tilde{E}$ & $C_6^-$ & ${\sigma}_{d3}$ & ${\sigma}_{d1}$ & ${\sigma}_{d2}$ & ${\sigma}_{v3}$ & ${\sigma}_{v1}$ & ${\sigma}_{v2}$ \\ \addlinespace[2pt]
        
        $\bs{C_3^-}$  & $C_3^-$ & $C_6^-$ & $\tilde{C}_2$ & $E$ & $\tilde{C}_3^+$ & $C_6^+$ & $\tilde{\sigma}_{d2}$ & $\tilde{\sigma}_{d3}$ & $\tilde{\sigma}_{d1}$ & $\tilde{\sigma}_{v2}$ & $\tilde{\sigma}_{v3}$ & $\tilde{\sigma}_{v1}$ & $\tilde{C}_3^-$ & $\tilde{C}_6^-$ & $C_2$ & $\tilde{E}$ & $C_3^+$ & $\tilde{C}_6^+$ & ${\sigma}_{d2}$ & ${\sigma}_{d3}$ & ${\sigma}_{d1}$ & ${\sigma}_{v2}$ & ${\sigma}_{v3}$ & ${\sigma}_{v1}$ \\ \addlinespace[2pt]
        
        $\bs{C_2}$  & $C_2$ & $\tilde{C}_3^-$ & $C_3^+$ & $\tilde{C}_6^-$ & $C_6^+$ & $\tilde{E}$ & ${\sigma}_{v1}$ & ${\sigma}_{v2}$ & ${\sigma}_{v3}$ & $\tilde{\sigma}_{d1}$ & $\tilde{\sigma}_{d2}$ & $\tilde{\sigma}_{d3}$ & $\tilde{C}_2$  & ${C}_3^-$ & $\tilde{C}_3^+$ & $C_6^-$ & $\tilde{C}_6^+$ & $E$ & $\tilde{\sigma}_{v1}$ & $\tilde{\sigma}_{v2}$ & $\tilde{\sigma}_{v3}$ & ${\sigma}_{d1}$ & ${\sigma}_{d2}$ & ${\sigma}_{d3}$ \\ \addlinespace[2pt]
        
        $\bs{\sigma_{d1}}$ & $\sigma_{d1}$ & $\sigma_{v3}$ & $\tilde{\sigma}_{v2}$ & $\tilde{\sigma}_{d2}$ & $\tilde{\sigma}_{d3}$ & $\tilde{\sigma}_{v1}$ & $\tilde{E}$ & $C_3^+$ & $C_3^-$ & $C_2$ & $C_6^-$ & $\tilde{C}_6^+$ & $\tilde{\sigma}_{d1}$ & $\tilde{\sigma}_{v3}$ & ${\sigma}_{v2}$ & ${\sigma}_{d2}$ & ${\sigma}_{d3}$ & $\sigma_{v1}$ & $E$ & $\tilde{C}_3^+$ & $\tilde{C}_3^-$ & $\tilde{C}_2$ & $\tilde{C}_6^-$ & $C_6^+$ \\ \addlinespace[2pt]
        
        $\bs{\sigma_{d2}}$ & $\sigma_{d2}$ & $\sigma_{v1}$ & $\tilde{\sigma}_{v3}$ & $\tilde{\sigma}_{d3}$ & $\tilde{\sigma}_{d1}$ & $\tilde{\sigma}_{v2}$ & $C_3^-$ & $\tilde{E}$ & $C_3^+$ & $\tilde{C}_6^+$ & $C_2$ & $C_6^-$ & $\tilde{\sigma}_{d2}$ & $\tilde{\sigma}_{v1}$ & ${\sigma}_{v3}$ & ${\sigma}_{d3}$ & $\sigma_{d1}$ & ${\sigma}_{v2}$ & $\tilde{C}_3^-$ & $E$ & $\tilde{C}_3^+$ & ${C}_6^+$ & $\tilde{C}_2$ & $\tilde{C}_6^-$ \\ \addlinespace[2pt]
        
        $\bs{\sigma_{d3}}$ & $\sigma_{d3}$ & $\sigma_{v2}$ & $\tilde{\sigma}_{v1}$ & $\tilde{\sigma}_{d1}$ & $\tilde{\sigma}_{d2}$ & $\tilde{\sigma}_{v3}$ & $C_3^+$ & $C_3^-$ & $\tilde{E}$ & $C_6^-$ & $\tilde{C}_6^+$ & $C_2$ & $\tilde{\sigma}_{d3}$ & $\tilde{\sigma}_{v2}$ & $\sigma_{v1}$ & $\sigma_{d1}$ & $\sigma_{d2}$ & $\sigma_{v3}$ & $\tilde{C}_3^+$ & $\tilde{C}_3^-$ & $E$ & $\tilde{C}_6^-$ & $C_6^+$ & $\tilde{C}_2$  \\ \addlinespace[2pt]
        
        $\bs{\sigma_{v1}}$ & $\sigma_{v1}$ & $\tilde{\sigma}_{d3}$ & $\sigma_{d2}$ & $\tilde{\sigma}_{v2}$ & $\tilde{\sigma}_{v3}$ & $\sigma_{d1}$ & $\tilde{C}_2$ & $\tilde{C}_6^-$ & $C_6^+$ & $\tilde{E}$ & $C_3^+$ & $C_3^-$ & $\tilde{\sigma}_{v1}$ & $\sigma_{d3}$ & $\tilde{\sigma}_{d2}$ & $\sigma_{v2}$ & $\sigma_{v3}$ & $\tilde{\sigma}_{d1}$ & $C_2$ & $C_6^-$ & $\tilde{C}_6^+$ & $E$ & $\tilde{C}_3^+$ & $\tilde{C}_3^-$  \\ \addlinespace[2pt]
        
        $\bs{\sigma_{v2}}$ & $\sigma_{v2}$ & $\tilde{\sigma}_{d1}$ & $\sigma_{d3}$ & $\tilde{\sigma}_{v3}$ & $\tilde{\sigma}_{v1}$ & $\sigma_{d2}$ & $C_6^+$ & $\tilde{C}_2$ & $\tilde{C}_6^-$ & $C_3^-$ & $\tilde{E}$ & $C_3^+$ & $\tilde{\sigma}_{v2}$ & $\sigma_{d1}$ & $\tilde{\sigma}_{d3}$ & $\sigma_{v3}$ & $\sigma_{v1}$ & $\tilde{\sigma}_{d2}$ & $\tilde{C}_6^+$ & $C_2$ & $C_6^-$ & $\tilde{C}_3^-$ & $E$ & $\tilde{C}_3^+$ \\ \addlinespace[2pt]
        
        $\bs{\sigma_{v3}}$ & $\sigma_{v3}$ & $\tilde{\sigma}_{d2}$ & $\sigma_{d1}$ & $\tilde{\sigma}_{v1}$ & $\tilde{\sigma}_{v2}$ & $\sigma_{d3}$ & $\tilde{C}_6^-$ & $C_6^+$ & $\tilde{C}_2$ & $C_3^+$ & $C_3^-$ & $\tilde{E}$ & $\tilde{\sigma}_{v3}$ & $\sigma_{d2}$ & $\tilde{\sigma}_{d1}$ & $\sigma_{v1}$ & $\sigma_{v2}$ & $\tilde{\sigma}_{d3}$ & $C_6^-$ & $\tilde{C}_6^+$ & $C_2$ & $\tilde{C}_3^+$ & $\tilde{C}_3^-$ & $E$ \\ \addlinespace[10pt]
        
        $\bs{\tilde{E}}$ & $\tilde{E}$ & $\tilde{C}_6^+$ & $\tilde{C}_6^-$ & $\tilde{C}_3^+$ & $\tilde{C}_3^-$ & $\tilde{C}_2$ & $\tilde{\sigma}_{d1}$ & $\tilde{\sigma}_{d2}$ & $\tilde{\sigma}_{d3}$ & $\tilde{\sigma}_{v1}$ & $\tilde{\sigma}_{v2}$ & $\tilde{\sigma}_{v3}$ & $E$ & $C_6^+$ & $C_6^-$ & $C_3^+$ & $C_3^-$ & $C_2$ & $\sigma_{d1}$ & $\sigma_{d2}$ & $\sigma_{d3}$ & $\sigma_{v1}$ & $\sigma_{v2}$ & $\sigma_{v3}$ \\ \addlinespace[2pt]
        
        $\bs{\tilde{C}_6^+}$ & $\tilde{C}_6^+$ & $\tilde{C}_3^+$ & $\tilde{E}$  & $\tilde{C}_2$ & $\tilde{C}_6^-$ & $C_3^-$ & $\sigma_{v2}$ & $\sigma_{v3}$ & $\sigma_{v1}$ & $\tilde{\sigma}_{d2}$ & $\tilde{\sigma}_{d3}$ & $\tilde{\sigma}_{d1}$ & $C_6^+$ & $C_3^+$ & $E$ & $C_2$ & $C_6^-$ & $\tilde{C}_3^-$ & $\tilde{\sigma}_{v2}$ & $\tilde{\sigma}_{v3}$ & $\tilde{\sigma}_{v1}$ & $\sigma_{d2}$ & $\sigma_{d3}$ & $\sigma_{d1}$ \\ \addlinespace[2pt]
        
        $\bs{\tilde{C}_6^-}$ & $\tilde{C}_6^-$ & $\tilde{E}$ & $\tilde{C}_3^-$ & $\tilde{C}_6^+$ & $C_2$ & $\tilde{C}_3^+$ & $\tilde{\sigma}_{v3}$ & $\tilde{\sigma}_{v1}$ & $\tilde{\sigma}_{v2}$ & $\sigma_{d3}$ & $\sigma_{d1}$ & $\sigma_{d2}$ & $C_6^-$ & $E$ & $C_3^-$ & $C_6^+$ & $\tilde{C}_2$ & $C_3^+$ & $\sigma_{v3}$ & $\sigma_{v1}$ & $\sigma_{v2}$ & $\tilde{\sigma}_{d3}$ & $\tilde{\sigma}_{d1}$ & $\tilde{\sigma}_{d2}$ \\ \addlinespace[2pt]
        
        $\bs{\tilde{C}_3^+}$ & $\tilde{C}_3^+$ & $\tilde{C}_2$ & $\tilde{C}_6^+$ & $C_3^-$ & $\tilde{E}$ & $C_6^-$ & ${\sigma}_{d3}$ & ${\sigma}_{d1}$ & ${\sigma}_{d2}$ & ${\sigma}_{v3}$ & ${\sigma}_{v1}$ & ${\sigma}_{v2}$ & $C_3^+$ & $C_2$ & $C_6^+$ & $\tilde{C}_3^-$ & $E$ & $\tilde{C}_6^-$ & $\tilde{\sigma}_{d3}$ & $\tilde{\sigma}_{d1}$ & $\tilde{\sigma}_{d2}$ & $\tilde{\sigma}_{v3}$ & $\tilde{\sigma}_{v1}$ & $\tilde{\sigma}_{v2}$ \\ \addlinespace[2pt]
        
        $\bs{\tilde{C}_3^-}$ & $\tilde{C}_3^-$ & $\tilde{C}_6^-$ & $C_2$ & $\tilde{E}$ & $C_3^+$ & $\tilde{C}_6^+$ & ${\sigma}_{d2}$ & ${\sigma}_{d3}$ & ${\sigma}_{d1}$ & ${\sigma}_{v2}$ & ${\sigma}_{v3}$ & ${\sigma}_{v1}$ & $C_3^- $ & $C_6^-$ & $\tilde{C}_2$ & $E$ & $\tilde{C}_3^+$ & $C_6^+$ & $\tilde{\sigma}_{d2}$ & $\tilde{\sigma}_{d3}$ & $\tilde{\sigma}_{d1}$ & $\tilde{\sigma}_{v2}$ & $\tilde{\sigma}_{v3}$ & $\tilde{\sigma}_{v1}$ \\ \addlinespace[2pt]
        
        $\bs{\tilde{C}_2}$ & $\tilde{C}_2$ & ${C}_3^-$ & $\tilde{C}_3^+$ & $C_6^-$ & $\tilde{C}_6^+$ & $E$ & $\tilde{\sigma}_{v1}$ & $\tilde{\sigma}_{v2}$ & $\tilde{\sigma}_{v3}$ & ${\sigma}_{d1}$ & ${\sigma}_{d2}$ & ${\sigma}_{d3}$ & $C_2$ & $\tilde{C}_3^-$ & $C_3^+$ & $\tilde{C}_6^-$ & $C_6^+$ & $\tilde{E}$ & ${\sigma}_{v1}$ & ${\sigma}_{v2}$ & ${\sigma}_{v3}$ & $\tilde{\sigma}_{d1}$ & $\tilde{\sigma}_{d2}$ & $\tilde{\sigma}_{d3}$\\ \addlinespace[2pt]
        
        $\bs{\tilde{\sigma}_{d1}}$ & $\tilde{\sigma}_{d1}$ & $\tilde{\sigma}_{v3}$ & ${\sigma}_{v2}$ & ${\sigma}_{d2}$ & ${\sigma}_{d3}$ & $\sigma_{v1}$ & $E$ & $\tilde{C}_3^+$ & $\tilde{C}_3^-$ & $\tilde{C}_2$ & $\tilde{C}_6^-$  & $C_6^+$ & $\sigma_{d1}$ & $\sigma_{v3}$ & $\tilde{\sigma}_{v2}$ &  $\tilde{\sigma}_{d2}$ & $\tilde{\sigma}_{d3}$ & $\tilde{\sigma}_{v1}$ & $\tilde{E}$ & $C_3^+$ & $C_3^-$ & $C_2$ & $C_6^-$ & $\tilde{C}_6^+$ \\ \addlinespace[2pt]
        
        $\bs{\tilde{\sigma}_{d2}}$ & $\tilde{\sigma}_{d2}$ & $\tilde{\sigma}_{v1}$ & ${\sigma}_{v3}$ & ${\sigma}_{d3}$ & $\sigma_{d1}$ & ${\sigma}_{v2}$ & $\tilde{C}_3^-$ & $E$ & $\tilde{C}_3^+$ & ${C}_6^+$ & $\tilde{C}_2$ & $\tilde{C}_6^-$ & $\sigma_{d2}$ & $\sigma_{v1}$ & $\tilde{\sigma}_{v3}$ & $\tilde{\sigma}_{d3}$ & $\tilde{\sigma}_{d1}$ & $\tilde{\sigma}_{v2}$ & $C_3^-$ & $\tilde{E}$ & $C_3^+$ & $\tilde{C}_6^+$ & $C_2$ & $C_6^-$ \\ \addlinespace[2pt]
        
        $\bs{\tilde{\sigma}_{d3}}$ & $\tilde{\sigma}_{d3}$ & $\tilde{\sigma}_{v2}$ & $\sigma_{v1}$ & $\sigma_{d1}$ & $\sigma_{d2}$ & $\sigma_{v3}$ & $\tilde{C}_3^+$ & $\tilde{C}_3^-$ & $E$ & $\tilde{C}_6^-$ & $C_6^+$ & $\tilde{C}_2$ & $\sigma_{d3}$ & $\sigma_{v2}$ & $\tilde{\sigma}_{v1}$ & $\tilde{\sigma}_{d1}$ & $\tilde{\sigma}_{d2}$ & $\tilde{\sigma}_{v3}$ & $C_3^+$ & $C_3^-$ & $\tilde{E}$ & $C_6^-$ & $\tilde{C}_6^+$ & $C_2$ \\ \addlinespace[2pt]
        
        $\bs{\tilde{\sigma}_{v1}}$ & $\tilde{\sigma}_{v1}$ & $\sigma_{d3}$ & $\tilde{\sigma}_{d2}$ & $\sigma_{v2}$ & $\sigma_{v3}$ & $\tilde{\sigma}_{d1}$ & $C_2$ & $C_6^-$ & $\tilde{C}_6^+$ & $E$ & $\tilde{C}_3^+$ & $\tilde{C}_3^-$ & $\sigma_{v1}$ & $\tilde{\sigma}_{d3}$ & $\sigma_{d2}$ & $\tilde{\sigma}_{v2}$ & $\tilde{\sigma}_{v3}$ & $\sigma_{d1}$ & $\tilde{C}_2$ & $\tilde{C}_6^-$ & $C_6^+$ & $\tilde{E}$ & $C_3^+$ & $C_3^-$ \\ \addlinespace[2pt]
        
        $\bs{\tilde{\sigma}_{v2}}$ & $\tilde{\sigma}_{v2}$ & $\sigma_{d1}$ & $\tilde{\sigma}_{d3}$ & $\sigma_{v3}$ & $\sigma_{v1}$ & $\tilde{\sigma}_{d2}$ & $\tilde{C}_6^+$ & $C_2$ & $C_6^-$ & $\tilde{C}_3^-$ & $E$ & $\tilde{C}_3^+$ & $\sigma_{v2}$ & $\tilde{\sigma}_{d1}$ & $\sigma_{d3}$ & $\tilde{\sigma}_{v3}$ & $\tilde{\sigma}_{v1}$ & $\sigma_{d2}$ & $C_6^+$ & $\tilde{C}_2$ & $\tilde{C}_6^-$ & $C_3^-$ & $\tilde{E}$ & $C_3^+$  \\ \addlinespace[2pt]
        
        $\bs{\tilde{\sigma}_{v3}}$ & $\tilde{\sigma}_{v3}$ & $\sigma_{d2}$ & $\tilde{\sigma}_{d1}$ & $\sigma_{v1}$ & $\sigma_{v2}$ & $\tilde{\sigma}_{d3}$ & $C_6^-$ & $\tilde{C}_6^+$ & $C_2$ & $\tilde{C}_3^+$ & $\tilde{C}_3^-$ & $E$ & $\sigma_{v3}$ & $\tilde{\sigma}_{d2}$ & $\sigma_{d1}$ & $\tilde{\sigma}_{v1}$ & $\tilde{\sigma}_{v2}$ & $\sigma_{d3}$ & $\tilde{C}_6^-$ & $C_6^+$ & $\tilde{C}_2$ & $C_3^+$ & $C_3^-$ & $\tilde{E}$ \\ \addlinespace[4pt]
        \bottomrule
\end{tabular}}\label{tab:c6vDGmultiplication}
\end{sidewaystable}

\end{document}